\def\draftversion{false}

\RequirePackage{ifthen}
\ifthenelse{\equal{\draftversion}{true}}{
  \documentclass[aps,prb,galley,showpacs,preprintnumbers,citeautoscript,
      amsmath,amssymb,longbibliography]{revtex4-1}
}{
  \documentclass[citeautoscript,floatfix,aps,prb,twocolumn,
      superscriptaddress,longbibliography]{revtex4-1}
}

\usepackage{amsmath}
\usepackage{graphicx}
\usepackage{epstopdf}
\usepackage{natbib}
\usepackage{array}
\usepackage{dcolumn}
\usepackage{bm}
\usepackage{multirow}
\usepackage{soul}  

\usepackage[usenames,dvipsnames]{color}

\newcommand\T{\rule{0pt}{2.6ex}}              
\newcommand\B{\rule[-1.2ex]{0pt}{0pt}}        

\soulregister\cite7

\ifthenelse{\equal{\draftversion}{true}}{
  \usepackage{showlabels}
  \marginparwidth 2.7in
  \marginparsep 0.5in
  \newcounter{comm} 
  \def\commnext{\stepcounter{comm}}
  \def\commtext{{\bf\color{blue}[\arabic{comm}]}}
  \def\commmar{{\bf\color{blue}[\arabic{comm}]}}
  \def\dvm#1{\commnext\marginpar{\small DV\commmar: #1}\commtext}
  \def\cdm#1{\commnext\marginpar{\small CED\commmar: #1}\commtext}
  \def\msm#1{\commnext\marginpar{\small MS\commmar: #1}\commtext}
  \def\asm#1{\commnext\marginpar{\small AS\commmar: #1}\commtext}
  \def\miq#1{\commnext\marginpar{\small MR\commmar: #1}\commtext}
  \def\mlab#1{\marginpar{\small\bf #1}}
  
}{
  \def\dvm#1{}
  \def\cdm#1{}
  \def\msm#1{}
  \def\asm#1{}
  \def\miq#1{}
  \def\mlab#1{}
  
}

\begin{document}

\title{Lattice-mediated bulk flexoelectricity from first principles}
\author{Miquel Royo}
\affiliation{Institut de Ci\`encia de Materials de Barcelona 
(ICMAB-CSIC), Campus UAB, 08193 Bellaterra, Spain}

\author{Massimiliano Stengel}
\affiliation{Institut de Ci\`encia de Materials de Barcelona 
(ICMAB-CSIC), Campus UAB, 08193 Bellaterra, Spain}
\affiliation{ICREA - Instituci\'o Catalana de Recerca i Estudis Avan\c{c}ats, 08010 Barcelona, Spain}

\date{\today}

\begin{abstract} 
We present the derivation and code implementation of a first-principles methodology to calculate the lattice-mediated 
contributions to the bulk flexoelectric tensor.
The approach is based on our recent analytical long-wavelength extension of 
density-functional perturbation theory [Royo and Stengel, Phys. Rev. X {\bf 9}, 021050 (2019)], 
and avoids the cumbersome numerical derivatives with respect to the wave vector that
were adopted in previous implementations.
To substantiate our results, we revisit and numerically validate the sum rules that relate 
flexoelectricity and uniform elasticity by generalizing them to regimes where finite forces and stresses 
are present.
We also revisit the definition of the elastic tensor under
stress, especially in regards to the existing linear-response
implementation.
We demonstrate the performance of our method by applying it to representative cubic crystals and to the tetragonal 
low-temperature polymorph of SrTiO$_3$, obtaining excellent agreement with the available literature data.

\end{abstract}

\maketitle

\section{Introduction}

Flexoelectricity, the physical property of insulators whereby a macroscopic polarization is induced by a strain gradient,
has received much attention in recent years.~\cite{pavlo_review,wang_review} Due to its universal character, it is a property of all dielectric materials regardless of crystal symmetry, which ideally offers low-cost and lead-free alternatives for typical piezoelectric materials operating in electromechanical devices.~\cite{zhu-06,Bhaskar2016}
Recent experimental work, either directly on nanomaterials~\cite{gustau1,noh_flexo,mcgilly-20} or by using the highly localized strain fields generated with a nanoscopic tip on a macroscopic surface,~\cite{lu-12,Yang-18,park-18,wang-20,park-20} has demonstrated the great potential of flexoelectricity at the nanoscale. 
Notable examples include the mechanical manipulation of the ferroelectric polarization;~\cite{gustau1,lu-12,park-18} the flexoelectronic effect, where a flexoelectric
voltage is used to gate transistor-like operations;~\cite{sharma-15,zhang-19,wang-20} or the flexo-photovoltaic effect, whereby a 
strain-gradient increases by orders of magnitude the photocurrent generated in a photovoltaic device.~\cite{Yang-18,Wu-21}

Future technologies based on flexoelectricity depend on maturing our understanding of the effect at the very fundamental level.
The first-principles theory and calculations of flexoelectricity have made impressive progress in recent years,
providing quantitatively predictive support to the interpretation of the experiments.
Achieving the current stage has taken nearly a decade of continuing 
efforts in order to settle the numerous formal subtleties.
Building on the classic method of the long-waves,~\cite{born/huang} 
the works of Resta,~\cite{resta-10} Hong and Vanderbilt,~\cite{hong-11,hong-13} and Stengel~\cite{artlin} have established
the general framework to define and compute the bulk response. Meanwhile, the specific role of the 
surfaces in finite samples was clarified~\cite{artgr} and incorporated in the formalism, enabling a 
pioneering application to SrTiO$_3$ slabs.~\cite{artcalc}
En route towards a practical implementation, additional technical and formal issues were also addressed, regarding, 
e.g. the treatment of inhomogeneous strains~\cite{StengelPRB18,SchiaffinoPRB19} and the proper definition of the 
current-density operator in the presence of nonlocal pseudopotentials.~\cite{cyrus}

Up to early 2019, first-principles calculations of flexoelectricity relied on numerical differentiation to extract the 
first- and second-order (in the wave vector {\bf q}) coefficients that define the flexoelectric tensor components 
within the long-wave approach.  
Such a procedure, however, is computationally cumbersome because of the necessity of performing
several linear-response calculations at small ${\bf q}$-vectors; moreover, exceedingly stringent convergence 
criteria are typically required for the perturbative expansion to be accurate.
Such limitations were overcome with the long-wave extension of the density-functional perturbation theory (DFPT) that we report in our preceding publication.~\cite{Royo2019} 
Long-wave DFPT enables an efficient calculation of a broad set of spatial-dispersion quantities, including those required to build the bulk flexoelectric tensor, thanks to an analytical formulation of the {\bf q}-derivatives.
In Ref.~\onlinecite{Royo2019}, we demonstrated the applicability of the method in the cases of the clamped-ion (CI) flexoelectric and dynamical quadrupole tensors, which are needed in order to build the  
electronic contributions to bulk flexoelectricity.

Here we present the theoretical formalism and computational implementation to calculate, within the long-wave DFPT framework, the 
remaining missing pieces to build the flexoelectric tensor at the bulk level. 
More specifically, we focus on the lattice-mediated contribution to the bulk flexoelectric tensor,
which we define and calculate in terms of two spatial-dispersion quantities: 
the first real-space moment of the zone-center force constants (FC) and the CI flexoelectric force-response 
tensor.~\cite{artlin,chapter-15} 
In addition, we revisit two sum rules, relating the latter two tensors to 
the uniform-strain response functions of Ref.~\onlinecite{hamann-metric}, and 
generalize them to crystals with finite (residual) forces and/or stresses.
Our developments have been incorporated among the
capabilities of an open-source public package (ABINIT v9.2~\cite{ABINIT2020}) and, in combination with the previously 
established electronic contributions,~\cite{Royo2019} allow a 
nonspecialized user to calculate the complete bulk flexoelectric tensor
of an arbitrary insulator, with the same computational efficiency that was 
demonstrated in Ref.~\onlinecite{Royo2019}.
To test and validate our implementation, we present calculations on
representative crystals, including semiconductors (Si,C) and perovskite-structure oxides
(SrTiO$_3$).
Note that an independent validation was already reported in Ref.~\onlinecite{springolo-21}, where
the present methods were applied to several 2D materials.
We shall also discuss a number of subtleties that arise in the physical interpretation
of the calculated flexoelectric coefficients.

This work is organized as follows.
In Sec.~\ref{sec:bulk_nu} we introduce the bulk flexoelectric tensor 
and the different intermediate quantities that enter its definition. 
We also recap the established sum rules that relate, at mechanical equilibrium, 
flexoelectricity to linear elasticity, and discuss the known ambiguities that plague 
the definition of of the bulk flexoelectric coefficients.
In Sec.~\ref{sec:lr}, we summarize 
the formalism of the long-wave DFPT (Ref.~\onlinecite{Royo2019}), 
and the metric-wave formulation of inhomogeneous strain perturbations 
(Ref.~\onlinecite{SchiaffinoPRB19}).
In Sec.~\ref{sec:lattice}, we derive the analytical formulas for the 
first real-space moment of the FC and the CI 
flexoelectric force-response tensor.
The sum rules that relate the latter two quantities to, respectively, 
the piezoelectric force-response tensor and the macroscopic elastic 
coefficients are generalized in Sec.~\ref{sec:elast} to a regime 
with finite atomic forces and stresses.
In Sec.~\ref{sec:results}, we demonstrate the performance and validity of our approach by 
applying it to several representative crystals, where we benchmark our results against 
the available literature data and numerically validate the revised sum rules. 
In Sec.~\ref{sec:conclusions}, we present our conclusions and outlook.
The Appendices provide additional analytic support for the formulas reported in the main text.

\section{Bulk flexoelectric tensor}

\label{sec:bulk_nu}

\subsection{Basic definitions}

\label{sec:basic}

To start with, we shall recap the first-principles theory of bulk flexoelectricity in its present state.
At difference with earlier works,~\cite{artlin,artgr,chapter-15} we adopt an alternative partition of the 
contributions entering the flexoelectric tensor, which we consider more symmetrical and conceptually appealing.
In close analogy with the theory of piezoelectricity,~\cite{martin} 
we split the bulk flexoelectric tensor into an electronic and a lattice-mediated contribution,
\begin{equation}
 \mu_{\alpha\gamma,\beta\delta}=\underbrace{\mu^{\rm el}_{\alpha\gamma,\beta\delta}}_{electronic} + \underbrace{\frac{1}{\Omega} Z^{(\alpha)}_{\kappa\rho} \widetilde{\Phi}^{(0)}_{\kappa\rho,\kappa'\sigma} C^{\kappa'}_{\sigma\gamma,\beta\delta}}_{lattice-mediated},
 \label{eq_fxetens}
\end{equation}
where $\kappa,\kappa'$ run over atomic sublattices, other indices refer to Cartesian directions, 
summation over repeated indexes is implied (here and in the rest of the manuscript), and the type-II form of the flexoelectric tensor has been implicitly adopted.~\cite{artlin,chapter-15}  
The lattice-mediated term is given by the product of the Born effective charges (BEC) tensor ($Z^{(\alpha)}_{\kappa\rho}$), the pseudoinverse of the zone-center FC matrix ($\widetilde{\Phi}^{(0)}_{\kappa\rho,\kappa'\sigma}$),~\cite{Wu-05} 
and the flexoelectric force-response tensor ($C^{\kappa'}_{\sigma\gamma,\beta\delta}$); 
$\Omega$ is the cell volume. 

At difference with the piezoelectric case, the electronic and force-response tensors of Eq.~(\ref{eq_fxetens}) 
are not elementary linear-response functions, but are further decomposed into a CI  and a remainder contribution, 
[CI quantities will be indicated by a bar hereafter]
\begin{subequations}
 \begin{align}
 \label{eq_ci_lr_elec}
  \mu^{\rm el}_{\alpha\gamma,\beta\delta} & =\bar{\mu}_{\alpha\gamma,\beta\delta} 
  - P_{\alpha,\kappa'\rho}^{(1,\gamma)} \Gamma^{\kappa'}_{\rho\beta\delta}, \\
 \label{eq_ci_lr_lat}
  C^{\kappa}_{\alpha\gamma,\beta\delta} & = \bar{C}^{\kappa}_{\alpha\gamma,\beta\delta} + \Phi^{(1,\gamma)}_{\kappa\alpha,\kappa'\rho} \Gamma^{\kappa'}_{\rho\beta\delta}.
 \end{align}
\label{eq_ci_lr}
\end{subequations}
Here, $\Gamma^{\kappa'}_{\rho\beta\delta}$ is the piezoelectric internal-strain tensor, describing the atomic displacements induced by a uniform strain, and is already available within the existing implementations~\cite{hamann-metric,Wu-05} of DFPT. 
The remaining four quantities are specific to flexoelectricity, and their calculation from first principles constitutes 
the main technical challenge.

The basic ingredients for their definition are the macroscopic electrical polarization (${\bf P}^{\bf q}_{\kappa \alpha}$) 
and force-constants matrix ($\Phi^{\bf q}_{\kappa \alpha,\kappa'\beta}$) that are associated with a phonon perturbation of the 
type
\begin{equation}
\label{phonon}
u^l_{\kappa \alpha} = \lambda e^{i{\bf q}\cdot {\bf R}_{l\kappa}},
\end{equation}
where $l$ is a cell index, $\lambda$ is the perturbation amplitude,
${\bf q}$ is the reciprocal-space momentum vector and ${\bf R}_{l\kappa}$ 
is the unperturbed atomic location. 
In the long-wavelength regime, we can expand the aforementioned quantities as~\cite{artlin}
\begin{subequations}
\label{taylor_lw}
\begin{align}
{\bf P}^{\bf q}_{\kappa \alpha} &\simeq {\bf P}^{(0)}_{\kappa \alpha} - iq_\gamma  {\bf P}^{(1,\gamma)}_{\kappa \alpha} 
 - \frac{q_\gamma q_\delta}{2} {\bf P}^{(2,\gamma\delta)}_{\kappa \alpha}, \\
\Phi^{\bf q}_{\kappa \alpha,\kappa'\rho} &\simeq \Phi^{(0)}_{\kappa\alpha,\kappa'\rho} -i q_\gamma \Phi^{(1,\gamma)}_{\kappa\alpha,\kappa'\rho} -
\frac{q_\gamma q_\delta}{2} \Phi^{(2,\gamma\delta)}_{\kappa\alpha,\kappa'\rho} .
\end{align}
\end{subequations}
$P_{\alpha,\kappa'\rho}^{(0)}$ and $\Phi^{(0)}_{\kappa\alpha,\kappa'\rho}$ correspond to the BEC tensor
and the zone-center FC, respectively, and are well established physical properties~\cite{Gonze/Lee,Baroni-01} 
in the framework of DFPT.
$P_{\alpha,\kappa'\rho}^{(1,\gamma)}$ and $\Phi^{(1,\gamma)}_{\kappa\alpha,\kappa'\rho}$, 
accounting for the second terms on the rhs of Eqs.~(\ref{eq_ci_lr}), can be regarded as the
first-order spatial dispersion counterparts of $P_{\alpha,\kappa'\rho}^{(0)}$ and $\Phi^{(0)}_{\kappa\alpha,\kappa'\rho}$ or, 
equivalently, as the first real-space moment of the microscopic polarization and forces produced by the displacement of 
an isolated atom.
The second-order expansion coefficients
in Eqs.~(\ref{taylor_lw}) are related to the first (CI) terms in Eqs.~(\ref{eq_ci_lr}) as follows. 
First, a simple summation over the sublattices yields the 
following type-I coefficients 
\begin{subequations}
\begin{align}
\bar{\mu}^{\rm I}_{\alpha\beta,\gamma\delta} &= \frac{1}{2}\sum_\kappa P^{(2,\gamma\delta)}_{\alpha, \kappa \beta} \\
 \left[\alpha\beta,\gamma\delta\right]^{\kappa} &=-\frac{1}{2}\sum_{\kappa'}\Phi^{(2,\gamma\delta)}_{\kappa\alpha,\kappa'\beta},
 \label{eq_sqrbkt} 
\end{align}
\end{subequations}
which are defined as the CI polarization and force response to a type-I strain gradient (second gradient of the
displacement field).
The following relation is then used to convert the result to type-II form (i.e., as a response to the first gradient of
the symmetric strain tensor),
\begin{subequations}
\begin{align}
\bar{\mu}_{\alpha\gamma,\beta\delta} &= \bar{\mu}^{\rm I}_{\alpha\beta, \gamma\delta} + \bar{\mu}^{\rm I}_{\alpha\delta,\beta \gamma} - \bar{\mu}^{\rm I}_{\alpha\gamma,\beta\delta}, \label{eq_mu_typeII}\\
\bar{C}^{\kappa}_{\alpha\gamma,\beta\delta} &= [\alpha\beta,\gamma\delta]^{\kappa} + [\alpha\delta,\beta\gamma]^{\kappa} -[\alpha\gamma,\beta\delta]^{\kappa}.
\label{eq_ci_fxfrt}
\end{align}
\end{subequations}

The present formulation differs from earlier works~\cite{artlin,chapter-15,ABINIT2020} 
in that the ``mixed'' contribution to the bulk flexoelectric
tensor has been reabsorbed here as part of the ``electronic'' response.
In addition to achieving a nicely symmetrical description of the ionic force and electronic polarization response 
to a strain gradient, this rearrangement is also desirable on general physical grounds. 
In a centrosymmetric crystal, the second terms on the rhs of
Eqs.~(\ref{eq_ci_lr}) are nonzero only in presence of ``gerade'' Raman-active modes (e.g., octahedral tilts in 
perovskites) that respond linearly to a uniform strain. (In absence of free Wyckoff parameters, the $\Gamma$ 
tensor identically vanishes.) 
A strain \emph{gradient} is associated with the gradients of such nonpolar modes
(their amplitude is determined by the local strain field);
the latter, in turn, couple to the macroscopic electronic polarization and 
to additional forces on the internal degrees of freedom via $P_{\alpha,\kappa'\rho}^{(1,\gamma)}$ 
and $\Phi^{(1,\gamma)}_{\kappa\alpha,\kappa'\rho}$, respectively.
Thus, the second terms in Eqs.~(\ref{eq_ci_lr}) can be understood physically as \emph{indirect}
contributions to the flexoelectric polarization 
that are mediated by gradients of the Raman-active modes.
Like direct flexoelectricity, such contributions consist in an electronic and lattice-mediated
part; the new decomposition of Eqs.~(\ref{eq_ci_lr}) correctly captures their respective 
physical nature, which was left implicit in earlier works.

\subsection{Sum rules at mechanical equilibrium \label{sec:sr}}

The physical meaning of the spatial-dispersion tensors entering Eq.~(\ref{eq_ci_lr_lat})
is best clarified via their relationship to known quantities within the theory of
linear elasticity. 
We shall recap the already established~\cite{born/huang,artlin} sum rules,
valid for a crystal at mechanical equilibrium (i.e., with vanishing internal forces
and stresses) in the following.

The first one relates the first moment of the FC with the piezoelectric force-response tensor,  
\begin{equation}
 \sum_{\kappa'} \Phi^{(1,\delta)}_{\kappa\alpha,\kappa'\beta}=\Lambda^{\kappa}_{\alpha\beta\delta},
 \label{eq_sr_pfr} 
\end{equation}
where the latter describes the forces induced on the sublattice $\kappa$ by a uniform strain $\eta_{\beta\delta}$.  
The product of $\Lambda^{\kappa}_{\alpha\beta\delta}$ with the pseudoinverse of the FC provides, in turn, the internal strain tensor,
\begin{equation}
 \Gamma^{\kappa}_{\alpha\beta\delta}=\widetilde{\Phi}^{(0)}_{\kappa\alpha,\kappa'\lambda} \Lambda^{\kappa'}_{\lambda\beta\delta},
 \label{eq_int_str}
\end{equation}
that appears in Eqs.~(\ref{eq_ci_lr}).

The second sum rule links the flexoelectric force-response tensor to the macroscopic elastic tensor 
$\mathcal{C}_{\alpha\gamma,\beta\delta}$ via~\cite{artlin}
\begin{equation}
 \frac{1}{\Omega}\sum_{\kappa} C^{\kappa}_{\alpha\gamma,\beta\delta}=\mathcal{C}_{\alpha\gamma,\beta\delta}.
 \label{eq_celast_ld}
\end{equation}
Note that $\mathcal{C}_{\alpha\gamma,\beta\delta}$ refers to the static elastic tensor,
and is customarily split into a CI contribution and another one due to internal relaxations 
of the ionic coordinates,~\cite{Wu-05}
\begin{equation}
 \mathcal{C}_{\alpha\gamma,\beta\delta}=\bar{\mathcal{C}}_{\alpha\gamma,\beta\delta} - \frac{1}{\Omega}
 \Lambda^{\kappa}_{\rho,\alpha\gamma} 
 \widetilde{\Phi}^{(0)}_{\kappa\rho,\kappa'\lambda} \Lambda^{\kappa'}_{\lambda,\beta\delta}.
 \label{eq_celast}
\end{equation}
By combining the two sum rules, Eq.~(\ref{eq_sr_pfr}) and (\ref{eq_celast_ld}), with Eq.~(\ref{eq_fxetens})
one can easily see that the elastic sum rule also holds at the CI level,
\begin{equation}
 \frac{1}{\Omega}\sum_{\kappa} \bar{C}^{\kappa}_{\alpha\gamma,\beta\delta}=\cal{\bar{C}}_{\alpha\gamma,\beta\delta}.
 \label{eq_sr_celast}
\end{equation}

The calculation of $\Lambda^{\kappa}_{\alpha\beta\delta}$ and $\cal{\bar{C}}_{\alpha\gamma,\beta\delta}$ is today commonly addressed with the metric-tensor approach of Hamann \emph{et al.}~\cite{hamann-metric} (HWRV), whereby they are obtained as DFPT second-order energy functionals for an atomic displacement plus a strain perturbations and for two strain perturbations, respectively.
In Sec.~\ref{sec:elast} we shall generalize Eqs.~(\ref{eq_sr_pfr}) and~(\ref{eq_sr_celast}) to the case of a crystal in presence of arbitrary forces and stresses. Then, in Sec.~\ref{sec:results} we shall validate our implementation by comparing the results of these generalized sum rules with the values directly obtained by means of HWRV linear-response calculations.

\subsection{Physical interpretation}

\label{sec:meaning}

The bulk flexoelectric coefficients defined in Eq.~(\ref{eq_fxetens}) and Eqs.~(\ref{eq_ci_lr}) 
should not be regarded as stand-alone physical properties of the crystal, but as intermediate quantities
that enable the calculation of ``proper'' experimental observables.
In particular, $\mu$ suffers from two distinct ambiguities, which we shall briefly summarize in 
the following.

The first issue concerns the imposition of short-circuit electrical boundary conditions (EBC),
which are a prerequisite~\cite{artlin} for taking the analytical long-wave expansions of 
Eq.~(\ref{taylor_lw}).
At the bulk level, short-circuit EBC are unambiguous at the zone center (in fact, 
they are automatically imposed by any electronic-structure code working in periodic boundary 
conditions), but the concept of ``macroscopic electric field''  becomes ill-defined when 
moving away from the ${\bf q}=0$ point.
As a consequence, the bulk flexoelectric coefficients are determined only modulo a constant term given by~\cite{artlin}
\begin{equation}
 \Delta\mu_{\alpha\gamma,\beta\delta}= \chi_{\alpha\gamma} \frac{\partial \mathcal{V}}{\partial \eta_{\beta\delta}},
 \label{eq_band_term}
\end{equation}
where $\chi_{\alpha\gamma}$ is the static dielectric susceptibility of the material;
$\mathcal{V}$ is an arbitrary (compatibly with crystal symmetries) scalar function of 
the cell parameters and atomic positions, with the physical dimension of a potential; and $\eta_{\beta\delta}$ 
is the uniform strain.
$\Delta\mu_{\alpha\gamma,\beta\delta}$ corresponds to the change 
in the bulk flexoelectric tensor produced by a shift of $\mathcal{V}$ in 
the reference potential that one uses to define the macroscopic electric 
field (and hence the short-circuit EBC).
All the individual physical properties entering Eqs.~(\ref{eq_ci_lr}), with the exception of the 
piezoelectric internal-strain tensor, are affected by this issue, which is usually referred to 
as a \emph{reference potential ambiguity}.~\cite{artlin} 
Following Ref.~\onlinecite{artlin}, we shall set the macroscopic electrostatic potential as the energy reference.

The second issue is specific to the lattice-mediated response, as defined in Eq.~(\ref{eq_fxetens}).
A consequence of Eq.~(\ref{eq_sr_celast}) is that, at difference with the piezoelectric force-response tensor ($\Lambda$), 
the sublattice sum of the flexoelectric force-response tensor ($C$) does not vanish.
Such a net force on the unit cell is problematic when multiplying $C^{\kappa}_{\alpha\gamma,\beta\delta}$
with the pseudoinverse in Eq.(1), as the resulting flexoelectric internal strains (and hence
the lattice-mediated polarization) will depend on the details of how the pseudoinverse is 
constructed.~\cite{hong-13, artlin,chapter-15} 
To prevent this, it is convenient to redefine 
the flexoelectric force-response tensor entering Eq.~(\ref{eq_fxetens}) 
as follows,
\begin{equation}
 \hat{C}^{\kappa}_{\alpha\gamma,\beta\delta}=C^{\kappa}_{\alpha\gamma,\beta\delta}
- \frac{w_{\kappa}}{\sum_{\kappa''} w_{\kappa''}} \sum_{\kappa'} C^{\kappa'}_{\alpha\gamma,\beta\delta},
\label{eq_mass_term}
\end{equation}
with $w_{\kappa}$ being an arbitrary set of positive weights. 
[One can also obtain the same result without 
explicitly using Eq.~(\ref{eq_mass_term}), but implicitly via a careful 
construction of the pseudoinverse.~\cite{hong-13}]
It is straightforward to verify that the sublattice
sum of $\hat{C}^{\kappa}_{\alpha\gamma,\beta\delta}$ vanishes regardless of the
specific choice of $w_{\kappa}$.
Following Ref.~\onlinecite{artlin}, we shall set the weights to the physical 
atomic masses in the present implementation. 
However, other choices are possible, as discussed e.g. in Refs.~\onlinecite{hong-13,artlin}. 

Note that the aforementioned ambiguities always cancel out when bulk flexoelectricity is
combined with other physical ingredients to predict some well-defined physical observables.
A classic example is the flexoelectric open-circuit potential in a bound sample, such as a 
capacitor or slab: the surface deformation potential suffers from the exact same 
reference potential indeterminacy as the bulk flexoelectric coefficient, but with the opposite sign,
yielding a total flexoelectric coefficient of the slab that is immune from the reference potential 
arbitrariness.
Similar considerations hold for the mass ambiguity, in relation to the controversial role
played by the ``dynamical flexoelectric effect''.~\cite{tagantsev,artlin} 
A detailed account of these issues can be found in the recent literature; in the remainder of
this paper we shall focus on the first-principles calculation of the bulk flexoelectric 
tensor within the above conventions for the reference potential and dynamical weights.

In summary, in order to obtain the bulk flexoelectric tensor, the ``new'' quantities we need 
to compute are the first and second moments of the FC matrix. 
As we shall illustrate shortly, both can be readily obtained by applying the long-wave formalism of Ref.~\onlinecite{Royo2019} to the second derivative of the total energy with respect to atomic displacements or acoustic phonons.~\cite{StengelPRB18,SchiaffinoPRB19}
In the following, 
we shall briefly summarize the main results of Ref.~\onlinecite{Royo2019} and Ref.~\onlinecite{SchiaffinoPRB19}.

\section{Linear-response techniques}

\label{sec:lr}

\subsection{Long-wave perturbation theory}
\label{sec:lw}

Consider two perturbations, $\lambda_1$ and $\lambda_2$, which are modulated at a
given wavevector ${\bf q}$.
The second derivatives of the total energy with respect to $\lambda_1$ and $\lambda_2$,
can be written, in the framework of DFPT, as a 
stationary (st) functional of the first-order wavefunctions plus a nonvariational (nv) 
contribution,
\begin{equation}
E^{\lambda_1^* \lambda_2}({\bf q}) = E_{\rm st}^{\lambda_1^* \lambda_2}({\bf q}) 
   + E_{\rm nv}^{\lambda_1^* \lambda_2}({\bf q}).
\label{e_tot}
\end{equation}
In full generality, the first part reads as
\begin{equation}
\begin{split}
& E_{\rm st}^{\lambda_1^* \lambda_2}({\bf q}) = 
  2s \int_{\rm {\rm BZ}} [d^3k] \sum_m E^{\lambda_1^* \lambda_2}_{m{\bf k}} ({\bf q}) \\
   & \quad + \int_{\Omega}\int K_{\bf{q}}({\bf r},{\bf r}') 
   n^{\lambda_1 *}_{\bf{q}}({\bf r}) 
   n^{\lambda_2}_{\bf{q}}({\bf r}') d^3r d^3r',
\label{e_va}
\end{split}
\end{equation}
where $n^{\lambda_{1,2}}_{\bf{q}}({\bf r})$ are the first-order electron densities, 
$K_{\bf{q}}({\bf r},{\bf r}')$ is the Coulomb and exchange-correlation kernel, and 
$s$ stands for the spin occupation factor. 
[Note that $E^{\lambda_1^* \lambda_2}$ is defined here as the second-order derivative 
of the total energy with respect to $\lambda_1$ and $\lambda_2$. 
This convention differs from that of Ref.~\onlinecite{Royo2019} and 
earlier works~\cite{Gonze-95a,gonze,gonze-97} 
by a factor of 2.]
The integrand in the first line, in turn, is a band-resolved contribution given by 
\begin{equation}
\begin{split}
 E^{\lambda_1^* \lambda_2}_{m{\bf k}} ({\bf q}) & = 
\langle u_{m{\bf k},{\bf q}}^{\lambda_1} | 
  \big( \hat{H}_{{\bf k}+{\bf q}}^{(0)} + a\hat{P}_{{\bf k}+{\bf q}} - \epsilon_{m{\bf k}} \big) 
   | u_{m{\bf k},{\bf q}}^{\lambda_2} \rangle \\
 & \quad + \langle u_{m{\bf k},{\bf q}}^{\lambda_1} | \hat{Q}_{{\bf k}+{\bf q}} 
   \hat{H}_{{\bf k},{\bf q}}^{\lambda_2} | u_{m{\bf k}}^{(0)} \rangle  \\
 & \quad + \langle u_{m{\bf k}}^{(0)} | \big( \hat{H}_{{\bf k},{\bf q}}^{\lambda_1} \big)^{\dagger} 
 \hat{Q}_{{\bf k}+{\bf q}} | u_{m{\bf k},{\bf q}}^{\lambda_2} \rangle,
 \label{e_band}
\end{split}
\end{equation}
where $| u_{m{\bf k}}^{(0)} \rangle$ and $| u_{m{\bf k},{\bf q}}^{\lambda_{1,2}} \rangle$ are,
respectively, the unperturbed and first-order Bloch states; $\hat{P}_{{\bf k}+{\bf q}}$ and
$\hat{Q}_{{\bf k}+{\bf q}} = 1 - \hat{P}_{{\bf k}+{\bf q}}$ are, respectively, the 
projectors on the valence- and conduction-band manifold;
$\hat{H}_{{\bf k},{\bf q}}^{\lambda_{1,2}}$ denotes the external perturbation
to the ground-state Hamiltonian, $\hat{H}_{{\bf k}}^{(0)}$; $\epsilon_{m{\bf k}}$
are the unperturbed Kohn-Sham eigenvalues; finally, $a$ is a constant of the dimension
of an energy that guarantees the unconstrained variational character of the 
functional.

The stationary character of $E_{\rm st}^{\lambda_1^* \lambda_2}({\bf q})$ allows for its 
perturbative long-wave expansion as a function of the parameter ${\bf q}$ without 
explicitly calculating the response to a gradient of the perturbations.
For example, in the notation of Ref.~\onlinecite{Royo2019} (a $\gamma$ subscript indicates derivation 
with respect to the wavevector component $q_\gamma$), at first order we have
\begin{equation}
\begin{split}
& E^{\lambda_1^* \lambda_2}_{\rm st,\gamma}= 2s \int_{{\rm BZ}} [d^3k] \sum_m E^{\lambda_1^* \lambda_2}_{m{\bf k},\gamma} \\
& \quad + \int_{\Omega}\int K_{\gamma}({\bf r},{\bf r}') n^{\lambda_1 *}({\bf r}) n^{\lambda_2}({\bf r}')d^3r d^3r',
\label{e_va1}
\end{split}
\end{equation}
with a band-resolved contribution
\begin{equation}
\begin{split}
& E^{\lambda_1^* \lambda_2}_{m{\bf k},\gamma} =
\langle u_{m{\bf k}}^{\lambda_1} | \hat{H}^{k_\gamma}_{{\bf k}} | u_{m{\bf k}}^{\lambda_2} \rangle \\
 & \quad - \sum_n \Big( \langle u_{m{\bf k}}^{\lambda_1} | u_{n{\bf k}}^{k_\gamma} \rangle
 \langle u_{n{\bf k}}^{(0)} |
 \hat{\mathcal{H}}_{{\bf k}}^{\lambda_2} | u_{m{\bf k}}^{(0)} \rangle \\
 & \qquad \quad \, \, + \langle u_{m{\bf k}}^{(0)} | 
 (\hat{\mathcal{H}}_{{\bf k}}^{\lambda_1})^{\dagger} | u_{n{\bf k}}^{(0)}  \rangle
   \langle u_{n{\bf k}}^{k_\gamma}  | u_{m{\bf k}}^{\lambda_2} \rangle \Big)
   \\
 & \quad + \langle u_{m{\bf k}}^{\lambda_1} | \hat{H}_{{\bf k},\gamma}^{\lambda_2} | u_{m{\bf k}}^{(0)} \rangle + 
 \langle u_{m{\bf k}}^{(0)} | (\hat{H}_{{\bf k},\gamma}^{\lambda_1})^{\dagger} | u_{m{\bf k}}^{\lambda_2} \rangle.
 \label{e_va_mk1}
\end{split}
\end{equation}
Here we have introduced the calligraphic symbol 
\begin{equation}
\hat{\mathcal{H}}_{{\bf k}}^{\lambda} = \hat{H}_{{\bf k}}^{\lambda} + V_{\bf k}^{\lambda}
\end{equation}
for the screened first-order Hamiltonian, inclusive of the self-consistent (SCF) Hartree and 
exchange-correlation potential, $V_{{\bf k}}^{\lambda}$; the superscript $k_\gamma$ indicates
the derivatives of the ground-state Hamiltonian and Bloch orbitals with respect to $k_\gamma$ 
($\hat{H}^{k_\gamma}_{{\bf k}}$ is the velocity operator); finally,
$K_{\gamma}$ stands for the {\bf q} gradient of the SCF kernel~\cite{Royo2019}. 

A set of second-order gradient formulas, slightly more complicated than the above ones, can also be derived when one of the perturbations $\lambda_1$ or $\lambda_2$ produces a vanishing response at $\bf{q}=0$.~\cite{Royo2019}
Both first- and second-order gradient functionals are fully general for any combination of perturbations $\lambda_{1,2}$.
They only require standard ingredients that are either already present within most 
implementations of DFPT (e.g., the first-order response functions to a phonon, electric field or
uniform strain), or can be analytically derived in a closed form (e.g., 
the ${\bf q}$-gradient of the external potentials, $\hat{H}_{{\bf k},\gamma}^{\lambda}$).
Explicit formulas for the latter were derived in Ref.~\onlinecite{Royo2019}, together
with practical examples for the two spatial-dispersion properties that enter the electronic flexoelectric tensor; their generalization to the lattice-mediated case is straightforward.
On the other hand, the nonvariational contribution to Eq.~(\ref{e_tot}) is specific to a given combination 
of $\lambda_1$ and $\lambda_2$; its derivation in the cases of relevance for this work will therefore constitute our main 
methodological task.

\subsection{The metric-wave perturbation}

\label{sec:metric}

The case of an acoustic phonon perturbation, where all sublattices are simultaneously
displaced with the same amplitude and phase, has a special significance
in the context of flexoelectricity.
Being this an elastic wave, it is convenient to treat it 
by operating a coordinate transformation to the curvilinear co-moving frame.
This way the atomic displacements are recast as a modulated change of the 
metric of space, which we shall indicate as a ``$(\beta)$'' symbol henceforth.

At the level of the first-order Hamiltonians, the following relationship holds
between the metric-wave (details can be found in Ref.~\onlinecite{SchiaffinoPRB19})
and the phonon perturbation of Eq.~(\ref{phonon}), indicated by $\tau_{\kappa\beta}$ for a displacement 
of the sublattice $\kappa$ along the Cartesian direction $\beta$ henceforth,
\begin{eqnarray}
\sum_\kappa \mathcal{\hat{H}}_{{\bf k},{\bf q}}^{\tau_{\kappa\beta}} 
  &=& \mathcal{\hat{H}}_{{\bf k},{\bf q}}^{(\beta)} + \Delta {\hat{H}}_{{\bf k},{\bf q}}^{\beta}, \nonumber
   \\
 \Delta {\hat{H}}_{{\bf k},{\bf q}}^{\beta} &=& i \hat{H}^{(0)}_{{\bf k}+{\bf q}} \Big(\hat{p}_{{\bf k}\beta} + \frac{q_{\beta}}{2} \Big) - i \Big(\hat{p}_{{\bf k}\beta} + \frac{q_{\beta}}{2} \Big) \hat{H}^{(0)}_{\bf k}, 
\label{eq_H_phon_vs_met_1} 
\end{eqnarray}
where $\hat{p}_{{\bf k}\beta}=-i \hat{\nabla}_{\beta} + k_{\beta}$ is the canonical momentum operator.
The following relationships then hold for the first-order wave functions
\begin{eqnarray}
 \sum_{\kappa}| u_{m{\bf k},{\bf q}}^{\tau_{\kappa\beta}} \rangle &=& | u_{m{\bf k},{\bf q}}^{(\beta)} \rangle + 
   | \Delta u_{m{\bf k},{\bf q}}^{\beta} \rangle \nonumber \\
 | \Delta u_{m{\bf k},{\bf q}}^{\beta} \rangle &=& 
-i \hat{Q}_{{\bf k}+{\bf q}} \Big(\hat{p}_{{\bf k}\beta} + \frac{q_{\beta}}{2} \Big) | u_{m{\bf k}}^{(0)} \rangle ,
\label{eq_wf1_phon_vs_met}
\end{eqnarray}
and densities,
\begin{eqnarray}
 \sum_{\kappa} n_{\bf q}^{\tau_{\kappa\beta}}({\bf r}) &=& n_{\bf q}^{(\beta)}({\bf r}) + \Delta n_{\bf q}^{\beta}({\bf r}) \nonumber \\
 \Delta n_{\bf q}^{\beta}({\bf r}) &=& - \frac{\partial n^{(0)}({\bf r})}{\partial r_{\beta}} - i q_{\beta} n^{(0)}({\bf r}).
 \label{eq_n_phon_vs_met}
\end{eqnarray}

The metric-wave Hamiltonian has two important properties that constitute a key advantage
in our context. First, the perturbation vanishes 
at ${\bf q}=0$,~\cite{SchiaffinoPRB19} 
\begin{equation}
\mathcal{\hat{H}}_{{\bf k},{\bf q}=0}^{(\beta)} = 0,
\label{metq0}
\end{equation}
as it should, since a rigid translation of the whole 
crystal has no effect in the comoving frame. 
Second, 
the first ${\bf q}$ gradient of the metric-wave perturbation directly
relates to the uniform strain formalism of HWRV,~\cite{hamann-metric}
\begin{equation}
 \hat{H}_{{\bf k},\delta}^{(\beta)}= i \hat{H}_{{\bf k}}^{\eta_{\beta\delta}}, \qquad 
 |u_{m{\bf k},\delta}^{(\beta)}\rangle= i |u_{m{\bf k}}^{\eta_{\beta\delta}}\rangle.
 \label{metq1}
\end{equation}

\section{Lattice-mediated contributions}

\label{sec:lattice}

\subsection{First order (phonon--phonon)}

Our starting point is the DFPT second-order energy response to two atomic displacements introduced by Gonze and Lee~\cite{gonze-97}. This functional allows one to obtain the FC matrix at any value of the wavevector {\bf q} as follows,
\begin{equation}
 \Phi_{\kappa\alpha,\kappa'\beta}({\bf q})= E^{\tau_{\kappa\alpha}^* \tau_{\kappa'\beta}}({\bf q}).
 \label{eq_Phi_def}
\end{equation}
The nonvariational term in this case consists in two contributions:
a ``geometric'' (ge) electronic term, written in 
terms of the ground-state wavefunctions, and the Ewald energy (Ew),
\begin{equation}
\begin{split}
E^{\tau^*_{\kappa\alpha}\tau_{\kappa'\beta}}_{\rm nv}({\bf q}) = 
  s \int_{{\rm BZ}} [d^3k]  E^{\tau^*_{\kappa\alpha}\tau_{\kappa'\beta}}_{{\rm ge},{\bf k}} 
& + E_{\rm{Ew}}^{\tau^*_{\kappa\alpha} \tau_{\kappa'\beta}}({\bf q}).
\label{eq_Phi_E2}
\end{split}
\end{equation}
The integrand is, in turn, written as a trace, 
\begin{equation}
 E^{\tau^*_{\kappa\alpha}\tau_{\kappa'\beta}}_{{\rm ge},{\bf k}} =  
 {\rm Tr} \, \left( \hat{H}^{\tau_{\kappa\alpha} \tau_{\kappa'\beta}}_{\bf k} \hat{P}_{\bf k} \right),
\end{equation}
where the second-order external potential is
\begin{equation}
\hat{H}^{\tau_{\kappa\alpha} \tau_{\kappa'\beta}}_{\bf k} = i \delta_{\kappa\kappa'} 
  \big[ \hat{H}_{\bf k}^{\tau_{\kappa\alpha}}, \hat{p}_{{\bf k}\beta} \big].
 \label{eq_Phi_E2mk}
\end{equation}
Note that the geometric term is independent of  {\bf q}, and therefore irrelevant to our scopes. 
Thus, we can write the first ${\bf q}$-gradient of the FC matrix as
\begin{equation}
 \Phi^{(1,\gamma)}_{\kappa\alpha,\kappa'\beta} = -{\rm Im} \, \left(  
   E_{{\rm st},\gamma}^{\tau^*_{\kappa\alpha} \tau_{\kappa'\beta}} + E_{\rm{Ew},\gamma}^{\tau^*_{\kappa\alpha} \tau_{\kappa'\beta}} \right),
 \label{eq_Phi1_def}
\end{equation}
where the second term in the round brackets is the first {\bf q} gradient of the ionic Ewald contribution. 
The latter (explicit formulas can be found in Appendix~\ref{app_ewald}) is actually the only ``new'' object 
entering the above functional, in the sense that it did not arise in the general spatial-dispersion formulas of our 
previous work.~\cite{Royo2019} 
 
\subsection{Second order (phonon--metric)}

\label{sec:ph-met}

In order to compute the square bracket tensor of Eq.~(\ref{eq_ci_fxfrt}), we need now to push
the long-wave expansion of the FC matrix to second order in ${\bf q}$.
This constitutes, at first sight, a major inconvenience: the full
$E_{{\rm st},\gamma\delta}^{\tau^*_{\kappa\alpha} \tau_{\kappa'\beta}}$ functional
requires, in principle, calculating the ${\bf q}$ gradient of the first-order wave functions, whose 
implementation would entail a substantial coding effort.
To avoid this, it suffices to observe that the full FC matrix is actually not needed -- 
the definition of the square brackets (Eq.~(\ref{eq_sqrbkt})) contains a summation over one of the sublattices.
Thus, in close analogy with the derivation of the CI flexoelectric tensor,~\cite{Royo2019} 
we recast the response as the second derivative of the energy with respect to a 
phonon and the metric wave,~\cite{SchiaffinoPRB19}
\begin{equation}
\sum_{\kappa'} E^{\tau_{\kappa\alpha}^* \tau_{\kappa'\beta}}({\bf q})= E^{\tau^*_{\kappa\alpha} (\beta)}({\bf q}).
\end{equation} 
We shall outline the formal procedure to do so in the following paragraphs.

First, by using the formal relationships between the metric-wave and
phonon perturbations that we have summarized in Sec.~\ref{sec:metric},
we rewrite the stationary part of the lhs as follows,
\begin{equation}
\begin{split}
\sum_{\kappa'}  E^{\tau_{\kappa\alpha}^* \tau_{\kappa'\beta}}_{\rm st}({\bf q})= &E^{\tau^*_{\kappa\alpha} (\beta)}_{\rm st}({\bf q}) +
s \int_{\rm {\rm BZ}} [d^3k]  \Delta E^{\tau^*_{\kappa\alpha} (\beta)}_{\bf k}({\bf q}), 
 \end{split}
\end{equation}
where $E^{\tau^*_{\kappa\alpha} (\beta)}_{\rm st}({\bf q})$ is consistently expressed in terms 
of $\hat{H}_{{\bf k},{\bf q}}^{(\beta)}$, $| u_{m{\bf k},{\bf q}}^{(\beta)} \rangle$ and $n_{\bf q}^{(\beta)}({\bf r})$
(together with the corresponding phonon counterparts) according to Eq.~(\ref{e_va}). 
The additional contribution originates from the second terms 
[$\Delta {\hat{H}}_{{\bf k},{\bf q}}^{\beta}$, $| \Delta u_{m{\bf k},{\bf q}}^{\beta} \rangle$ and 
$\Delta n_{\bf q}^{\beta}({\bf r})$] on the rhs of Eqs.~(\ref{eq_H_phon_vs_met_1})--(\ref{eq_n_phon_vs_met}),
\begin{equation}
\begin{split}
 \Delta & E^{\tau^*_{\kappa\alpha} (\beta)}_{\bf k}({\bf q}) = \\ &  -2i \sum_m \langle u_{m{\bf k}}^{(0)} | 
 \big( \hat{H}_{{\bf k},{\bf q}}^{\tau_{\kappa \alpha}} \big)^{\dagger} 
\hat{Q}_{{\bf k}+{\bf q}} \Big(\hat{p}_{{\bf k}\beta} + \frac{q_{\beta}}{2} \Big) | u_{m{\bf k}}^{(0)} \rangle.
\label{deltae}
\end{split}
\end{equation}

Crucially, Eq.~(\ref{deltae}) only contains ground-state wavefunctions and operators,
and therefore can be readily reabsorbed in the geometric term,
\begin{equation}
E_{{\rm ge},{\bf k}}^{\tau_{\kappa\alpha}(\beta)} ({\bf q}) = \sum_{\kappa'}  E^{\tau_{\kappa\alpha}^* \tau_{\kappa'\beta}}_{\rm ge,\bf k} +
  \Delta E^{\tau^*_{\kappa\alpha} (\beta)}_{\bf k}({\bf q}).
\end{equation}  
Eq.~(\ref{deltae}), in turn, can be further simplified 
by writing $\hat{Q}_{{\bf k}+{\bf q}}=1-\hat{P}_{{\bf k}+{\bf q}}$ and by
observing that the contribution of the valence band projector vanishes after performing the implicit band 
summation and Brillouin zone integration. 
[This point was demonstrated in Ref.~\onlinecite{SchiaffinoPRB19} for the first order density 
response to a metric wave.] 
This implies that the geometric contribution can still be written as a trace,
\begin{equation}
 E_{{\rm ge},{\bf k}}^{\tau_{\kappa\alpha}(\beta)} ({\bf q})= 
 {\rm Tr} \, \left( \hat{H}^{\tau_{\kappa\alpha}(\beta)}_{\bf k, q} \hat{P}_{\bf k} \right),
 \label{eq_geom0}
\end{equation} 
where the new second-order Hamiltonian is
\begin{equation}
\begin{split}
 \hat{H}^{\tau_{\kappa\alpha}(\beta)}_{\bf k, q} = & i [ \hat{H}^{\tau_{\kappa\alpha}}_{\bf k} , \hat{p}_{{\bf k}\beta} ] 
  -2 i \big( \hat{H}_{{\bf k},{\bf q}}^{\tau_{\kappa\alpha}} \big)^{\dagger} \Big( \hat{p}_{{\bf k}\beta} + \frac{q_{\beta}}{2} \Big).
\end{split}
\end{equation}  
[The first term on the rhs comes from a trivial sublattice summation of Eq.~(\ref{eq_Phi_E2mk}), while
the second term embodies the contribution of Eq.~(\ref{deltae}).]

Finally, the Ewald contribution is simply written as the sublattice sum of its 
phonon counterpart,
\begin{equation}
E_{\rm{Ew}}^{\tau^*_{\kappa\alpha} (\beta)} ({\bf q})= 
\sum_{\kappa'} E_{\rm{Ew}}^{\tau^*_{\kappa\alpha} \tau_{\kappa'\beta}}({\bf q}).
\end{equation}
Toghether with the results of the above paragraphs, this allows us to write the second-order energy 
at finite ${\bf q}$ in the same form of Eq.~(\ref{e_tot}) and Eq.~(\ref{eq_Phi_E2}),
\begin{equation}
E^{\tau^*_{\kappa\alpha} (\beta)}({\bf q}) = E_{\rm st}^{\tau^*_{\kappa\alpha} (\beta)}({\bf q}) + 
E_{\rm nv}^{\tau^*_{\kappa\alpha} (\beta)}({\bf q}).
\label{E2_phon_met}
\end{equation}
\begin{equation}
\begin{split}
E^{\tau^*_{\kappa\alpha}(\beta)}_{\rm nv}({\bf q}) = 
  s \int_{{\rm BZ}} [d^3k]  E^{\tau^*_{\kappa\alpha}(\beta)}_{{\rm ge},{\bf k}} ({\bf q}) 
& + E_{\rm{Ew}}^{\tau^*_{\kappa\alpha} (\beta)}({\bf q}).
\label{eq_atdismet_nv}
\end{split}
\end{equation}
One can easily verify that, at ${\bf q}=0$, the acoustic sum rule on the force response to a rigid translation
is exactly satisfied, since the stationary, geometric and Ewald contributions all vanish individually.

The two properties of the metric wave, Eq.~(\ref{metq0}) and Eq.~(\ref{metq1}), 
allow us to push the long-wave expansion to second order in ${\bf q}$, and obtain
the square-bracket tensor as 
\begin{equation}
  [\alpha \beta,\gamma \delta]^\kappa =   \frac{1}{2} E^{\tau^*_{\kappa\alpha} (\beta)}_{\gamma \delta}.
\label{eq_srb_func}
\end{equation}
The second derivative in ${\bf q}$ is carried out separately for
the stationary, geometric and Ewald contributions.
The stationary term is straightforward to differentiate in ${\bf q}$
by following the prescriptions of Ref.~\onlinecite{Royo2019}: the 
procedure is exactly the same as in the
case of the CI flexoelectric tensor. 
[Further details on the derivation are reported in Appendix~\ref{sec:E_st}.]
Regarding the second gradient of the geometric contribution, 
 the second-order Hamiltonian is expanded as follows,
\begin{equation}
  \hat{H}^{\tau_{\kappa\alpha}(\beta)}_{\bf k, \gamma \delta} = 
   -i \Big[ 2(\hat{H}_{{\bf k },\gamma\delta}^{\tau_{\kappa\alpha}})^{\dagger}  \hat{p}_{{\bf k }\beta} 
 + (\hat{H}_{{\bf k },\gamma}^{\tau_{\kappa\alpha}})^{\dagger}  \delta_{\beta\delta}
 +(\hat{H}_{{\bf k },\delta}^{\tau_{\kappa\alpha}})^{\dagger}  \delta_{\beta\gamma} \Big] .
 \label{eq_frozen}
\end{equation}
Explicit formulas for the second gradient of the 
atomic displacement Hamiltonian ($\hat{H}_{{\bf k },\gamma\delta}^{\tau_{\kappa\alpha}}$) 
and of the Ewald term are reported in Appendix~\ref{app_Hatdis_d2dqdq} and Appendix~\ref{app_ewald},
respectively.

\section{Relationship to elasticity}

\label{sec:elast}

The two sum rules that we discussed in Sec.~\ref{sec:bulk_nu} establish a direct 
relationship between the present theory of flexoelectricity and some
known quantities in the context of elasticity. 
These results, which can be directly related to the 
classic treatment by Born and Huang,~\cite{born/huang} are strictly valid 
for a crystal at mechanical equilibrium; such an assumption
has always been adopted in earlier works.
Our present implementation is more general, though: the calculation
of the flexoelectric force-response tensor and the spatial dispersion
of the FC can be carried out successfully in any crystal, 
even in presence of arbitrary forces and stresses.
Our goal in this Section will be to revise Eqs.~(\ref{eq_sr_pfr}) and~(\ref{eq_sr_celast}) in order to take such a possibility into account.
Our derivations are based on the theory of Barron and Klein,~\cite{Barron-65}
which we shall revisit within a modern linear-response context.

\subsection{Unsymmetrized strain tensor}

The link to elasticity is provided by the definition of the unsymmetrized 
strain, ${\bf u}$, as a linear transformation of the lattice of the 
following form,
\begin{equation}
\label{unsymm}
{\bf R}'_{l\kappa} = \underbrace{({\bf I} + {\bf u})}_{\bf h} {\bf R}_{l\kappa},
\end{equation}
where ${\bf R}'_{l\kappa}$ and ${\bf R}_{l\kappa}$ are the perturbed and
unperturbed atomic locations, respectively, ${\bf u}$ is a $3\times 3$
tensor, and ${\bf I}$ is the identity matrix.
The quantity in the round bracket is known as \emph{deformation gradient},
and is defined, within a transformation ${\bf r} \rightarrow {\bf r}'$, as 
the Jacobian
\begin{equation}
h_{\alpha \beta}= \frac{\partial r'_\alpha}{\partial r_\beta};
\end{equation}
then the unsymmetrized strain ${\bf u}$ is simply
the gradient of the displacement field, where the
latter is defined as ${\bf r}'-{\bf r}$.

With these definitions, we can write down an expansion of the energy with respect to 
lattice-periodic distortions $\tau_{\kappa \alpha}$ and unsymmetrized strains ${\bf u}$ of the following type,
\begin{equation}
\begin{split}
E-E_0=& - f_{\kappa \alpha} \tau_{\kappa \alpha} + \Omega_0 S_{\alpha \gamma} u_{\alpha \gamma} 
 + \frac{1}{2} \Phi_{\kappa \alpha,\kappa' \beta} \tau_{\kappa \alpha} \tau_{\kappa' \beta}\\ 
 &-\frac{\partial f_{\kappa \alpha}}{\partial u_{\beta \delta}}\tau_{\kappa\alpha} u_{\beta \delta}
+ \frac{\Omega_0}{2} \bar{C}^{\rm un}_{\alpha \gamma, \beta \delta} u_{\alpha \gamma} u_{\beta \delta} + \cdots.
\label{eofu}
\end{split}
\end{equation}
Here $\Omega_0$ is the cell volume in the reference configuration;
 $f_{\kappa \alpha}$ are the atomic forces; $S_{\alpha \gamma}$ is the stress tensor, symmetric in $(\alpha,\gamma)$; and 
we have defined the \emph{unsymmetrized-strain elastic tensor} as 
\begin{equation}
\bar{C}^{\rm un}_{\alpha \gamma,\beta\delta} = \frac{1}{\Omega_0} \frac{\partial^2 E}{\partial u_{\alpha \gamma} \partial u_{\beta \delta}}.
\label{def_imp_celast}
\end{equation}
[We shall, in the following, exclusively focus on the sum rule Eq.~(\ref{eq_sr_celast}),
involving the elastic tensor components at the CI level.]

To establish a connection between the above and the quantities
that we define and calculate in this work, we shall rewrite Eq.~(\ref{eofu})
in reciprocal space [an alternative proof, formulated in real space, is reported in Appendix~\ref{app_realspace}], where the displacement field $u_\alpha$ is written in the same form as Eq.~(\ref{phonon}),
\begin{equation}
u^l_{\kappa \alpha} = u_\alpha e^{i{\bf q}\cdot{\bf R}^l_\kappa}.
\end{equation}
(All sublattices participate to the distortion with the same 
amplitude $u_\alpha$, times a phase factor that depends on the
location of the atom at rest.)
The unsymmetrized strain at a given point is then given by
\begin{equation}
u_{\alpha \beta}({\bf r}) = i u_\alpha q_\beta e^{i{\bf q}\cdot{\bf r}}.
\end{equation}
This allows us to write the second-order part of Eq.~(\ref{eofu})
in the following form,
\begin{equation}
\begin{split}
E-E_0=&  \frac{1}{2} \Phi_{\kappa \alpha,\kappa' \beta} \tau_{\kappa \alpha} \tau_{\kappa' \beta}\\ 
 &-i q_\delta \frac{\partial f_{\kappa \alpha}}{\partial u_{\beta \delta}}\tau_{\kappa\alpha} u_{\beta}
+ q_\gamma q_\delta \frac{\Omega_0}{2} \bar{C}^{\rm un}_{\alpha \gamma, \beta \delta} u_{\alpha} u_{\beta} + \cdots.
\label{eofq}
\end{split}
\end{equation}
We can now compare this result with the corresponding expression that emerges from Eq.~(\ref{taylor_lw}),~\cite{sge}
\begin{equation}
\begin{split}
E-E_0=&  \frac{1}{2} \Big[ \Phi^{(0)}_{\kappa\alpha,\kappa'\beta }\tau_{\kappa \alpha} \tau_{\kappa' \beta} \\
 &  -i q_\gamma \sum_{\kappa'} \Phi^{(1,\gamma)}_{\kappa\alpha,\kappa'\beta} \tau_{\kappa \alpha} u_\beta +
     i q_\gamma \sum_{\kappa} \Phi^{(1,\gamma)}_{\kappa\alpha,\kappa'\beta} u_\alpha \tau_{\kappa' \beta}   \\
 & - \frac{q_\gamma q_\delta}{2} \sum_{\kappa \kappa'} \Phi^{(2,\gamma\delta)}_{\kappa\alpha,\kappa'\beta} u_\alpha u_\beta   \Big].
\label{eofq2}
\end{split}
\end{equation}
By recalling the definition of the square brackets, Eq.~(\ref{eq_sqrbkt}),
we arrive at the following sum rules
\begin{subequations}
\begin{align}
\sum_{\kappa'} \Phi_{\kappa \alpha,\kappa' \beta}^{(1,\delta)}        = & \frac{\partial f_{\kappa \alpha}}{\partial u_{\beta \delta}}, \\
\frac{1}{\Omega}\sum_\kappa [\alpha\beta,\gamma\delta]^{\kappa} = & \frac{1}{2} (\bar{C}^{\rm un}_{\alpha \gamma,\beta\delta} + 
       \bar{C}^{\rm un}_{\alpha \delta, \beta\gamma}),
\end{align}
\end{subequations}
relating the quantities defined in this work to the internal force and macroscopic stress response
to the unsymmetrized strain perturbation described by Eq.~(\ref{unsymm}).

Note that, if the sample is stressed in the initial configuration, $\bf{\bar{C}^{\rm un}}$ does not 
satisfy all the symmetries under exchanges of indices that one expects for
the elastic tensor.
Of course, its definition as a second derivative ensures 
invariance with respect to the exchange of the first and second set of indices;
the additional symmetry with respect to $(\alpha,\gamma)$ or $(\beta, \delta)$ is, however, not guaranteed.
Thus, the above results for the sum rules are not yet satisfactory; other
definitions of the elastic tensor are preferrable under stress, as we shall see in
the following.

\subsection{Lagrange strain}

The Lagrange strain tensor (also known as ``finite-strain'' tensor) is
defined in terms of the deformation gradient, $h_{\alpha \beta} = \delta_{\alpha \beta} + u_{\alpha \beta}$,
as $\bm{\varepsilon} = ({\bf h}^{\rm T} {\bf h} - {\bf I})/2$. 
It can be equivalently expressed in terms of the unsymmetrized strain tensor as
\begin{equation}
\varepsilon_{\alpha \beta} = \frac{1}{2} \left(u_{\alpha \beta} + u_{\beta \alpha} + u_{\gamma \alpha}u_{\gamma \beta} \right).
\label{eta}
\end{equation}
Note that this expression reduces to the standard definition of the symmetrized strain tensor
within a regime of small deformations (Cauchy's theory).
Lagrange's formalism has, however, a crucial advantage, in that it expresses the 
macroscopic strain via the \emph{metric tensor} of the deformation, ${\bf g} = {\bf h}^{\rm T} {\bf h}$;
this automatically guarantees invariance of the theory with respect to arbitrary rotations
of the reference frame.
Also, formulating macroscopic elasticity in terms of the metric tensor of space naturally 
leads~\cite{hamann-metric} to a
lattice (reduced) representation of the internal coordinates, which are related
to the Cartesian positions via 
\begin{equation}
\tau_{\kappa \alpha} = h_{\alpha \beta} \hat{\tau}_{\kappa \beta} = (\delta_{\alpha\beta} + u_{\alpha \beta}) \hat{\tau}_{\kappa \beta}.
\end{equation}
The above definitions lead to an expansion of the total energy per unit cell that is analogous
to Eq.~(\ref{eofu}),
\begin{equation}
\begin{split}
E-E_0=& - f_{\kappa \alpha} \hat{\tau}_{\kappa \alpha} + \Omega_0 S_{\alpha \gamma} \varepsilon_{\alpha \gamma} 
 + \frac{1}{2} \Phi_{\kappa \alpha,\kappa' \beta} \hat{\tau}_{\kappa \alpha} \hat{\tau}_{\kappa' \beta}\\ 
 &-\Lambda^{\kappa}_{{\alpha \beta \delta}} \hat{\tau}_{\kappa\alpha} \varepsilon_{\beta \delta}
+ \frac{\Omega_0}{2} \bar{C}^{\rm La}_{\alpha \gamma, \beta \delta} \varepsilon_{\alpha \gamma} \varepsilon_{\beta \delta} + \cdots,
\label{eofe}
\end{split}
\end{equation}
The \emph{Lagrange elastic tensor} $\bar{C}^{\rm La}_{\alpha \gamma, \beta \delta}$ has the required symmetries: 
$(\alpha \gamma, \beta \delta)$, $(\alpha,\gamma)$ and $(\beta, \delta)$, a property 
that originates from the symmetry of $\varepsilon_{\alpha\gamma}$.

At first order, the expansion coefficients are the same; however, they are
multiplied by Cartesian displacements and unsymmetrized strain in the case
of Eq.~(\ref{eofu}), and by reduced-coordinate displacements and Lagrange
strain in Eq.~(\ref{eofe}).
More specifically, we have
\begin{subequations}
\begin{align}
f_{\kappa \alpha} \tau_{\kappa \alpha} = & f_{\kappa \alpha} \hat{\tau}_{\kappa \alpha} + f_{\kappa \alpha} u_{\alpha \beta}
\hat{\tau}_{\kappa \beta}, \\
S_{\alpha \gamma} \varepsilon_{\alpha \gamma} = &  S_{\alpha \gamma} u_{\alpha \gamma} + \frac{1}{2} S_{\alpha \gamma} u_{\delta \alpha}u_{\delta \gamma}.
\end{align}
\end{subequations}
By equating the relevant second-order terms of Eq.~(\ref{eofu}) and Eq.~(\ref{eofe}),
we obtain the following result
\begin{eqnarray}
\frac{\partial f_{\kappa \alpha}}{\partial u_{\beta \delta}} + f_{\kappa \beta} \delta_{\alpha \delta} &=& 
     \Lambda^{\kappa}_{{\alpha \beta \delta}}, \\
\label{eq_imp_celast}
\bar{C}^{\rm un}_{\alpha \gamma, \beta \delta} &=& \delta_{\alpha \beta} S_{\gamma \delta} + \bar{C}^{\rm La}_{\alpha \gamma, \beta \delta}.
\end{eqnarray}
Finally, by using this result we can write down our sum rules as follows,
\begin{eqnarray}
\label{eq_gsr_pfr}
\sum_{\kappa'} \Phi_{\kappa \alpha,\kappa' \beta}^{(1,\delta)}  &=& 
         \Lambda^{\kappa}_{{\alpha \beta \delta}} - f_{\kappa \beta} \delta_{\alpha \delta}, \\
\label{eq_gsr_celast}
\frac{1}{\Omega} \sum_\kappa [\alpha \beta,\gamma \delta]^{\kappa} &=& \frac{1}{2} (\bar{C}^{\rm La}_{\alpha \gamma, \beta \delta} + \bar{C}^{\rm La}_{\alpha \delta, \beta \gamma}) + \delta_{\alpha \beta} S_{\gamma \delta}.
\label{bh2}
\end{eqnarray}
The last equation can be inverted to obtain an expression, written in terms of the flexoelectric force-response tensor, that generalizes Eq.~(\ref{eq_sr_celast}),
\begin{equation}
 \frac{1}{\Omega}\sum_{\kappa} \bar{C}^{\kappa}_{\alpha\gamma,\beta\delta}= 
 \bar{C}^{\rm La}_{\alpha\gamma,\beta\delta} 
 + \delta_{\alpha\beta} S_{\gamma\delta}
 + \delta_{\alpha\delta} S_{\beta\gamma}
 - \delta_{\alpha\gamma} S_{\beta\delta}.
 \label{eq_gsr_celast_pract}
\end{equation}

\subsection{Hamann's linear-response approach}

We are only left to discuss the above results in the context of the existing implementations
of the strain perturbation within linear-response theory, most notably that of HWRV.~\cite{hamann-metric}
It is straightforward to show that HWRV's code implementation of the piezoelectric 
force-response tensor corresponds to our definition of $\Lambda^{\kappa}_{{\alpha \beta \gamma}}$ given in Eq.~(\ref{eq_gsr_pfr}). 
The situation regarding the elastic tensor is a bit more 
confused. Eq.~(31) of Ref.~\onlinecite{hamann-metric} describes it as the 
second derivative of the energy with respect to the unsymmetrized strain,
i.e., as the tensor ${\bf \bar{C}^{\rm un}}$ defined above. 
However, this cannot correspond to the actual implementation:
As we said, ${\bf \bar{C}^{\rm un}}$ in general does not comply with the 
required Voigt symmetry of the first and second pair of indices.

In an unpublished document (available on the ABINIT website~\cite{og}),
A.R. Oganov attempts to clarify this issue by suggesting that the 
first derivative of the stress tensor is implemented instead.
In our notation, his definition reads as
\begin{equation}
\bar{C}_{\alpha \gamma,\beta\delta}^{\rm Og} = \frac{1}{\Omega_0} 
 \frac{\partial}{\partial u_{\beta\delta}} (\Omega \sigma_{\alpha \gamma}),
\end{equation} 
where $\Omega$ and $\sigma_{\alpha \gamma}$ are, respectively, the cell volume and stress after the deformation $u_{\beta\delta}$.
[The strain derivative is intended to be taken at the CI level, consistent with the earlier paragraphs of this Section.]
Based on the definitions of the above paragraphs, and following 
the arguments of Barron and Klein,~\cite{Barron-65} we can readily express $\bar{C}_{\alpha \gamma,\beta\delta}^{\rm Og}$
in terms of the Lagrange elastic tensor as
\begin{equation}
\label{oganov}
\bar{C}_{\alpha \gamma,\beta\delta}^{\rm Og} = \bar{C}_{\alpha\gamma,\beta\delta}^{\rm La} + 
\delta_{\beta\gamma} S_{\alpha \delta}  + \delta_{\alpha\beta}S_{\gamma \delta} .
\end{equation}
This expression shows that $\bar{C}_{\alpha \gamma,\beta\delta}^{\rm Og}$ 
generally violates the symmetry with respect to $(\beta,\delta)$, and also
the symmetry with respect to $(\alpha \gamma,\beta\delta)$.
For this reason, Eq.~(\ref{oganov}) can hardly be regarded as a 
definition of the ``proper'' elastic tensor, contrary to Oganov's 
claims; clearly, Eq.~(\ref{oganov}) does not match the existing ABINIT
implementation, either.
After a detailed analysis of the latter, we conclude that HWRV's 
linear-response code calculates a tensor that is related to 
$\bar{C}_{\alpha \gamma,\beta\delta}^{\rm Og}$ by a symmetrization 
with respect to the last pair of indices,
\begin{equation}
\begin{split}
& \mathcal{\bar{C}}^{\rm HWRV}_{\alpha\gamma,\beta\delta}= 
\frac{\bar{C}_{\alpha \gamma,\beta\delta}^{\rm Og} + \bar{C}_{\alpha\gamma ,\delta\beta}^{\rm Og}}{2}=
\bar{C}_{\alpha\gamma,\beta\delta}^{\rm La} +  \\
&  \quad \frac{1}{2} \left( \delta_{\gamma\beta} S_{\alpha\delta}
+ \delta_{\alpha\beta} S_{\gamma\delta}  + \delta_{\alpha\delta}S_{\gamma\beta}  + \delta_{\gamma\delta}
S_{\alpha\beta}  \right).
\label{hwrv_celast}
\end{split}
\end{equation}
One can quickly verify that the resulting tensor now complies with 
the required symmetries.
We shall use HWRV's method, in conjunction with Eq.~(\ref{hwrv_celast}) in 
our numerical tests of the sum rules.

\begin{figure}
\begin{center}
\includegraphics[width=3.2in]{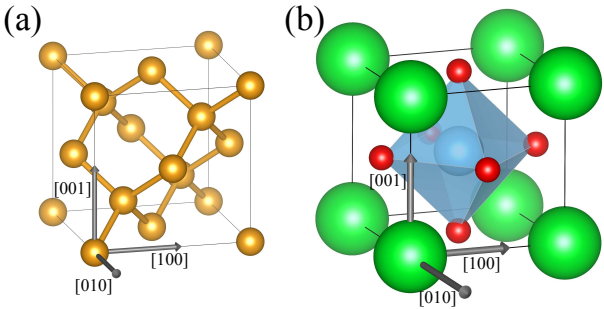}
\caption{\label{fig_cubic} Unit cell of (a) the diamond-structure and (b) the c-SrTiO$_3$ perovskite simulated materials. Grey arrows illustrate the crystallographic directions that form the Cartesian axes.}
\end{center}
\end{figure}

The above derivation, in addition to settling the existing formal 
issues regarding HWRV's method, also allows for a more compact 
expression of our second sum rule.
To that end, note that the strain derivative of the stress is related
to Oganov's definition via~\cite{og,hamann-metric}
\begin{equation}
\frac{\partial \sigma_{\alpha \gamma}}{\partial u_{\beta \delta}} = \bar{C}_{\alpha \gamma,\beta\delta}^{\rm Og}
   - \delta_{\beta \delta} S_{\alpha \gamma} .
\end{equation}
By combining this result with Eq.~(\ref{oganov}), we obtain an
equivalent formulation of Eq.~(\ref{eq_gsr_celast_pract}) as
\begin{equation}
 \frac{1}{\Omega}\sum_{\kappa} \bar{C}^{\kappa}_{\alpha\gamma,\beta\delta}= \bar{\cal{C}}_{\alpha\gamma,\beta\delta}=
  \frac{\partial \sigma_{\beta \delta} }{\partial u_{\alpha \gamma}}.
 \end{equation}

The strain derivative of the stress tensor is indicated by HWRV as ``improper'' elastic tensor. 
Here we have seen that there are many flavors of the elastic tensor
so the distinction between what should be regarded as a ``proper'' and ``improper''
definition needs to be revised in light of the above results. 
We shall call ``proper'' any definition that respects the full symmetries of the elastic
tensor under the most general (anisotropic stress) conditions. 
Only ${\bf C^{\rm La}}$ and 
${\bf C^{\rm HWRV}}$ fulfill this requirement, although we regard ${\bf C^{\rm La}}$ as a physically more
useful choice. [It lends itself more easily to higher-order generalizations of
the elastic energy, see e.g. Ref.~\onlinecite{cao2018}.] 
The other definitions, ${\bf C^{\rm un}}$, ${\bf C^{\rm Og}}$ and $\bm{\mathcal{C}}$, are all ``improper'' in the above sense.

\begin{table*}
\setlength{\tabcolsep}{9.0pt}
\begin{center}
\caption{Breakdown of the linearly independent bulk flexoelectric coefficients (in nC/m) of  Si, diamond (C) and c-SrTiO$_3$. The first four columns show the contributions arising from the different terms in Eqs.~(\ref{eq_ci_lr}). The total flexoelectric tensor is shown in the fifth column. The rightmost column shows the open-circuit flexovoltages (in V). Highlighted quantities have been directly obtained with the formalism and implementation presented in this work.}
\begin{tabular}{rc|rrrrr|r}\hline\hline
 \T \B &  & \multicolumn{1}{c}{$\bm{\bar{\mu}}$} & \multicolumn{1}{c}{$\bm{P}^{(1)}\bm{\Gamma}$} &  \multicolumn{1}{c}{$\bm{\bar{C}}$} & \multicolumn{1}{c}{$\bm{\Phi}^{(1)}\bm{\Gamma}$} & \multicolumn{1}{c}{$\bm{\mu}$} & \multicolumn{1}{c}{$\bm{\varphi}$} \\\hline
\T\B & $xx,xx$ & $-$1.399 &        &    &  & $-$1.399 & $-$12.002 \\
Si \T\B & $xx,yy$ & $-$1.036 &        &    &  & $-$1.036 & $-$8.894 \\
\T\B & $xy,xy$ & $-$0.188 & {\bf $-$0.107}  &    &   & $-$0.296 & $-$2.538\\ \hline 
\T\B & $xx,xx$  & $-$0.995  &        &    &  & $-$0.995 & $-$19.683\\
C \T\B & $xx,yy$ & $-$0.787 &        &    &  & $-$0.787 & $-$15.567 \\
\T\B & $xy,xy$ & $-$0.131 & {\bf $-$0.009}  &    &  &$-$0.141 & $-$2.783\\ \hline
\T\B & $xx,xx$ & $-$0.891  &   &  {\bf $-$181.950}  &  & $-$182.841 & $-$18.583\\
SrTiO$_3$ \T\B & $xx,yy$ & $-$0.832  &   &  {\bf $-$157.953} &  & $-$158.785 & $-$16.138\\
\T\B & $xy,xy$ & $-$0.083  &   &  {\bf $-$19.310}   &  & $-$19.392 & $-$1.971 \\ \hline \hline 
\label{Tab_cubic_flexo}
\end{tabular}
\end{center}
\end{table*}

 \section{Implementation and results}
 
 \label{sec:results}
 
 \begin{figure}
\begin{center}
\includegraphics[width=3.2in]{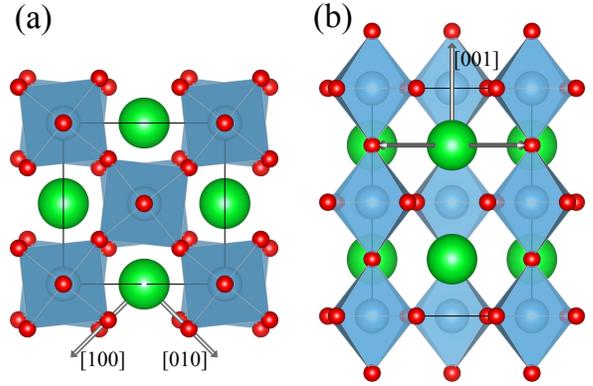}
\caption{\label{fig_tetr} (a) Top and (b) side view of the unit cell employed in the simulation of t-SrTiO$_3$. Grey arrows illustrate the crystallographic directions that form the Cartesian axes.}
\end{center}
\end{figure}

  \subsection{Implementation details \label{sec:imp_det}}
 
  The calculation of the tensors $\Phi^{(1,\gamma)}_{\kappa\alpha,\kappa'\beta}$ and  
  $\bar{C}^\kappa_{\alpha\gamma,\beta\delta}$ by 
  means of Eqs.~(\ref{eq_Phi1_def}) and~(\ref{eq_Csupk}), respectively, has been implemented in the \textsc{abinit} open-source software.~\cite{Gonze2009,Gonze2016,ABINIT2020} In addition, the post-processing computations in order to obtain the  indirect contributions to the flexoelectric tensor (i.e., the rhs terms of Eqs.~(\ref{eq_ci_lr})) have been implemented in the \textsc{anaddb} tool. These developments, together with the ones associated with our preceding work,~\cite{Royo2019} provide a new functionality, recently released for its public usage in the v9 version of \textsc{abinit},~\cite{ABINIT2020} that allows a non-specialized user to readily obtain the complete bulk flexoelectric tensor for any time-reversal symmetric crystalline insulator.
  
  The internal strain $\Gamma^{\kappa}_{\rho\beta\delta}$ is necessary to build the indirect terms of Eqs.~(\ref{eq_ci_lr}). 
  As we have seen in Sec.~\ref{sec:bulk_nu}, its calculation first requires to compute the piezoelectric force-response tensor, 
  $\Lambda_{\alpha,\beta\delta}^{\kappa}$.
  The latter, in turn, can be directly obtained from a uniform strain linear-response calculation~\cite{hamann-metric}, 
  as implemented in the \textsc{abinit} package, or from the sublattice summation of the first moment of the FC matrix 
  via Eq.~(\ref{eq_gsr_pfr}). 
  In our implementation we have adopted the second option for consistency with the other quantities entering the 
  flexoelectric tensor. 
  In any case, we test numerically their mutual relationship, Eq.~(\ref{eq_gsr_pfr}), to validate our implementation
  of $\Phi^{(1,\gamma)}_{\kappa\alpha,\kappa'\beta}$. 
  To calculate the \emph{improper} contributions of Eqs.~(\ref{eq_gsr_pfr}) and~(\ref{eq_gsr_celast}) we use
  the standard~\cite{NielsenPRB85} formulation of forces and stress as implemented in \textsc{abinit}.
  The pseudoinverse of the FC matrix $\widetilde{\Phi}^{(0)}_{\kappa\rho,\kappa'\sigma}$ is constructed following the prescriptions of Ref.~\onlinecite{Wu-05}.

   \subsection{Computational parameters}

 Our numerical results are obtained using the DFT and DFPT implementations of \textsc{abinit} v9.0.4 with the Perdew-Wang~\cite{perdew/wang:1992} parametrization of the LDA. We use norm-conserving pseudopotentials from the Pseudo Dojo,~\cite{pseudodojo} after regenerating them without exchange-correlation nonlinear core corrections using the ONCVPSP software.~\cite{Hamann1979,Hamann1989} 
 The unit cells employed for our simulations on cubic- and tetragonal-structure materials are illustrated in Figs.~\ref{fig_cubic} and~\ref{fig_tetr},
 respectively.
 
 For our calculations on Si and diamond, we use the primitive 2-atom cells with a cell parameter of 10.182 bohr (Si) and 6.673 bohr (diamond), a plane-wave cutoff of 20 Ha (Si) and 40 Ha (diamond) and a BZ sampled with a Monkhorst-Pack (MP) mesh of $12 \times 12 \times 12$ {\bf k} points. 
 For our calculations of cubic SrTiO$_3$ we use a 5-atom primitive cell, with a lattice constant of $a=7.288$ bohr,
  a plane-wave cutoff of 80 Ha and a BZ sampling of $10\times10\times10$  {\bf k} points.
  We represent the tetragonal structure with a 20-atom cell (cell parameters: $a=10.271$, $c=14.639$ bohr).
  To facilitate the comparison to the cubic structure we align the pseudocubic 
  [100], [010] and [001] directions (the latter corresponding to the tilt axis) 
  with the Cartesian $x$, $y$ and $z$ axes, respectively. 
 We use a BZ sampling of ($6\times6\times4$) {\bf k} points in this case.

\begin{table}[b!]
\setlength{\tabcolsep}{6pt}
\begin{center}
\caption{\label{Tab_diam_mix} List of parameters driving the indirect contribution to flexoelectricity in the diamond structure: first moment of the FC ($\phi$, units are $10^{-3}$ Ha/bohr), piezoelectric force-response from a HWRV calculation~\cite{hamann-metric} ($\lambda$, $10^{-3}$ Ha/bohr), frequency of the Raman-active mode ($\omega_{T_{1g}}$, cm$^{-1}$), internal strain ($\gamma$, $10^{-3}$ bohr) and the electronic polarization induced by an atomic-displacement gradient ($p$, $10^{-3}$ e/bohr$^{2}$).} 

\begin{tabular}{c|rrrrr}\hline\hline
   \T\B  & $\phi$ & \multicolumn{1}{c}{$\lambda$} & \multicolumn{1}{c}{$\omega_{T_{1g}}$} &  \multicolumn{1}{c}{$\gamma$}  &  \multicolumn{1}{c}{$p$} \\ 
   \hline
	Si \T\B & 190.272 & 190.273 &  512.555 & 681.428 & 25.961 \\
	C \T\B  &  79.060 &  79.071  &  1323.381 & 99.315 & 15.355 \\
  \hline \hline
\end{tabular}
\end{center}
\end{table}

 \subsection{Cubic crystals \label{sec:cubic}}
 
 We start by testing the performance of our implementation on three representative 
 insulators with cubic crystal structure: Si, diamond and SrTiO$_3$. 
 The main motivation originates from the availability of first-principles results
 in the literature on 
 both the first moment of the FC matrix and the CI flexoelectric force-response tensor. 
 [For details on the purely electronic response functions contributing to the 
 flexoelectric tensor, i.e., $\bar{\mu}_{\alpha\gamma,\beta\delta}$ and $P_{\alpha,\kappa'\rho}^{(1,\gamma)}$  
 in Eq.~(\ref{eq_ci_lr_elec}), see Ref.~\onlinecite{Royo2019}.]
 Note that, due to cubic symmetry, the flexoelectric tensor in this materials set
 has only three independent components; following earlier works, we shall refer to them as
 longitudinal ($\mu_{xx,xx}$), transverse ($\mu_{xx,yy}$) and shear ($\mu_{xy,xy}$).
 In Table~\ref{Tab_cubic_flexo} we present a summary of the results for the three
 materials; we provide their detailed analysis hereafter.

 \subsubsection{Diamond-structure semiconductors}

\begin{table}
\setlength{\tabcolsep}{8.0pt}
\begin{center}
\caption{Linearly-independent components of the silicon and diamond CI flexoelectric force-response tensors (in eV) and of the CI elastic tensors (in GPa). The latter are calculated with the sublattice summation of the flexoelectric force-response tensor (Eq.~(\ref{eq_sr_celast})) and with the HWRV approach.~\cite{hamann-metric} }
\centering
\begin{tabular}{rc|rrr}\hline\hline
 \T \B &  & \multicolumn{1}{r}{$xx,xx$} & \multicolumn{1}{r}{$xx,yy$} &  \multicolumn{1}{r}{$xy,xy$} \\\hline
   \T\B & $\bar{C}^{\kappa}_{\alpha\gamma,\beta\delta}$   & 19.670  & 7.678  & 12.880 \\
Si \T \B & $\mathcal{\bar{C}}_{\alpha\gamma,\beta\delta}$ & 161.163 & 62.903  & 105.532 \\
   \T\B & $\mathcal{\bar{C}}^{\rm HWRV}_{\alpha\gamma,\beta\delta}$ & 161.169 & 62.907  & 105.533 \\ \hline
   \T\B & $\bar{C}^{\kappa}_{\alpha\gamma,\beta\delta}$   & 38.148  & 4.999   & 20.712    \\
C  \T \B & $\mathcal{\bar{C}}_{\alpha\gamma,\beta\delta}$ & 1110.651 & 145.560 & 603.005 \\  
   \T\B & $\mathcal{\bar{C}}^{\rm HWRV}_{\alpha\gamma,\beta\delta}$ & 1110.899 & 145.903  & 602.955   \\
    \hline \hline 
\label{Tab_diam_forces}
\end{tabular}
\end{center}
\end{table}

Due to their purely covalent nature, the Born effective charge tensor identically vanishes in both diamond and Si,
which implies that the lattice-mediated contribution to the flexoelectric tensor vanishes as well.
The purely electronic part, Eq.~(\ref{eq_ci_lr_elec}), consists of a clamped-ion contribution~\cite{Royo2019} 
and an additional term originated from the strain coupling of the Raman-active optical mode;
we shall focus on the latter in the following. 
Via a symmetry analysis, we find that the relevant tensors are governed by one parameter each,
\begin{subequations}
\label{eqs_indir}
\begin{align}
P_{\alpha,\kappa\rho}^{(1,\gamma)} &=&  (-1)^{\kappa+1}\,p\,|\varepsilon_{\alpha\rho\gamma}|, \\
\Phi^{(1,\delta)}_{\kappa\rho,\kappa'\beta}&=&(-1)^{\kappa+1}\, (1-\delta_{\kappa\kappa'}) \, \phi \,|\varepsilon_{\rho\beta\delta}|, \label{diam_phi1}\\
\Lambda^{\kappa}_{\rho\beta\delta} &=& (-1)^{\kappa+1}\,\lambda \,|\varepsilon_{\rho\beta\delta}|, \label{diam_lambda}\\
\Gamma^{\kappa}_{\rho\beta\delta} &=& (-1)^{\kappa+1}\,\gamma \,|\varepsilon_{\rho\beta\delta}|,
\end{align}
\end{subequations}
where $\varepsilon_{\alpha\beta\gamma}$ is the Levi-Civita tensor. 
The calculated values reported in Table~\ref{Tab_diam_mix} show that the piezoelectric 
force-response tensor sum rule at mechanical equilibrium, Eq.~(\ref{eq_sr_pfr}), 
is fulfilled to a very high degree of accuracy in both materials.
[Due to Eqs.~(\ref{diam_phi1}) and~(\ref{diam_lambda}), this sum rule trivially 
reduces to an equality: $\phi=\lambda$.]
Our results for $\gamma$ are in fair agreement with the value of $\gamma= 0.72,0.14$ bohr, respectively
for Si and C, quoted by Hong and Vanderbilt~\cite{hong-13}. 
Our values of $p$, however,
markedly differ with those of Ref.~\onlinecite{hong-13}; this is likely due to a technical issue with 
the calculations reported therein~\cite{david_pc}.
Note that the $p$ coefficients in both Si and C 
are related to the dynamical quadrupoles $Q$ of Ref.~\onlinecite{Royo2019} via $p= Q /(2 \Omega)$.

Based on Eq.~(\ref{eq_ci_lr_elec}), the indirect (``mixed'') contribution to the
flexoelectric tensor in Si and C can be written as
\begin{equation}
\mu^{\rm ind}_{\alpha\gamma,\beta\delta} = -2p\gamma |\delta_{\alpha \beta} \delta_{\gamma \delta} -  
\delta_{\alpha \delta} \delta_{\gamma \beta}|,
\end{equation}
yielding a value of $\mu^{\rm ind}_{\alpha\beta,\alpha\beta} =\mu^{\rm ind}_{\alpha\beta,\beta\alpha} = 
 -2p\gamma$ for the shear components and zero otherwise.

In Table~\ref{Tab_cubic_flexo} we summarize the results
for the bulk flexoelectric tensor in Si and C.
The values shown 
indicate that $\mu^{\rm ind}$ is 
significantly smaller than the purely electronic contribution $\bar\mu$, especially in C.
[$\mu^{\rm ind}$ accounts for $\sim 37\%$ and  $\sim 6\%$ of the total shear coefficient in Si and C, respectively.]
We ascribe the smallness of the response in diamond to the stiffness of the Raman mode (see Table~\ref{Tab_diam_mix}), 
which leads to a strongly suppressed internal-strain parameter $\Gamma$ compared to Si. 
The electronic polarization parameter $P$,
on the other hand, appears comparable in both materials.

Even if the force-response tensor does not result in a contribution to the flexoelectric 
polarization, 
it is still interesting to analyze it in light of the sum rules discussed in Sec.~\ref{sec:sr}.
We find that, at the CI level, $\bar{C}^{\kappa}_{\alpha\gamma,\beta\delta}$ is symmetric 
with respect to the atomic index $\kappa$; thus, Eq.~(\ref{eq_sr_celast}) reduces again (within
the zero-stress regime that we test here) to an equality,
$\bar{\mathcal{C}} = 2\bar{C}^{\kappa=1} / \Omega$.
The values reported in Table~\ref{Tab_diam_forces} show that this sum rule
holds to an excellent degree of accuracy, confirming the correctness and numerical 
quality of our implementation.

\begin{table}
\setlength{\tabcolsep}{5pt}
\begin{center}
\caption{\label{Tab_STO_forces} Five top rows: Linearly-independent CI flexoelectric force-response tensor coefficients, $\bar{C}^{\kappa}_{\alpha\gamma,\beta\delta}$, of c-SrTiO$_3$  (in eV). Data between brackets is from Ref.~\onlinecite{chapter-15}. Two bottom rows: CI elastic coefficients (in GPa) calculated with the sublattice summation of the flexoelectric force-response tensor (Eq.~(\ref{eq_sr_celast})) and with the HWRV approach~\cite{hamann-metric}.  } 
\begin{tabular}{c|rrr}\hline\hline
  Atom \T\B  & \multicolumn{1}{c}{$xx,xx$} & \multicolumn{1}{c}{$xx,yy$} &  \multicolumn{1}{c}{$xy,xy$} \\\hline
	Sr \T\B & $-$25.0 ($-$24.9)  &  $-$28.5 ($-$28.7)  & 7.7 (7.9) \\
	Ti \T\B & $-$66.9 ($-$67.9) &  $-$102.2 ($-$102.3)  & 3.5 (3.8) \\
	O1 \T\B & 34.9 (35.2) &  30.7 (30.9)  & 17.2 (17.3) \\
	O2 \T\B & 34.9 (35.2) &  42.3 (42.3)  & 15.3 (15.3) \\
	O3 \T\B & 159.9 (159.3) &  97.4 (97.4)  & $-$1.0 ($-$0.9) \\ \hline
	$\mathcal{\bar{C}}_{\alpha\gamma,\beta\delta}$ \T\B & \multicolumn{1}{c}{384.842} & \multicolumn{1}{c}{110.680} & \multicolumn{1}{c}{119.041} \\
	$\mathcal{\bar{C}}^{\rm HWRV}_{\alpha\gamma,\beta\delta}$ \T\B &  \multicolumn{1}{c}{384.848} & \multicolumn{1}{c}{110.671} & \multicolumn{1}{c}{119.039}\\
	\hline \hline 
\end{tabular}
\end{center}
\end{table}

 \subsubsection{Cubic SrTiO$_3$}

In cubic SrTiO$_3$ (c-SrTiO$_3$),
the absence of free Wyckoff parameters results in vanishing
${\bm P^{(1)}}$ and ${\bm \Phi^{(1)}}$ tensors; 
this implies that the indirect contributions to the flexoelectric polarization vanish as well.
The most striking feature emerging from our results (Table~\ref{Tab_cubic_flexo}) is that 
the lattice-mediated contribution largely dominates the overall response in this material, accounting for 
more than $99\%$ of the total in all three tensorial components.
This is consistent with our expectations and the results of earlier works~\cite{artcalc,sge}: cubic SrTiO$_3$, being an incipient ferroelectric, 
has a polar mode (of $T_{1u}$ symmetry) with low frequency (45.738 cm$^{-1}$) and strong dipolar character. This mode  
mediates a huge electrical response to any symmetry-allowed perturbation (for example, the static dielectric constant 
nearly diverges in SrTiO$_3$ at low temperature), dominating over other degrees of freedom.
In the experimental context, in ferroelectrics and incipient ferroelectrics
it has become common practice to divide the measured flexoelectric coefficients
by the static dielectric constant, and thus obtain a temperature-independent 
physical quantity known as \emph{flexovoltage}. 
For this reason, we also report these coefficients ($\varphi$) in Table~\ref{Tab_cubic_flexo}, in units of voltage.

Consistent with earlier works, the longitudinal and transverse components of the
flexoelectric tensor (Table~\ref{Tab_cubic_flexo}) are similar in magnitude, and
both much larger than the shear component.
This is largely a consequence of using the macroscopic electrostatic potential as 
the (arbitrary) energy reference when imposing short-circuit EBC (see e.g. Sec.~\ref{sec:meaning}),
and should not be interpreted as a real physical effect.
Indeed, for the case of c-SrTiO$_3$, it was shown~\cite{artgr} that all three types of 
strain-gradient deformations yield a quantitatively similar flexoelectric response when 
other energy references, such as the valence- and conduction-band edges, are adopted.
Discussing the details on how the choice of the energy reference affects each individual 
contribution appearing in  Eqs.~(\ref{eq_ci_lr}) goes beyond the scopes of this work,
and will be covered in a future publication.

\begin{table}
\setlength{\tabcolsep}{10pt}
\begin{center}
\caption{\label{Tab_dist_pfr} Selected piezoelectric force-response coefficients (in $10^{-3}$ Ha/bohr) of the distorted SrTiO$_{3}$ system. The first column shows results from a HWRV~\cite{hamann-metric} metric-tensor calculation, while the second and third columns show the sublattice summation of the first moment of the FC matrix neglecting (Eq.~(\ref{eq_sr_pfr})) and including (Eq.~(\ref{eq_gsr_pfr})) the improper contribution mentioned in the text.} 
\begin{tabular}{c|rrr}\hline\hline
  \T\B  & \multicolumn{1}{c}{HWRV} &  \multicolumn{1}{c}{Eq.~(\ref{eq_sr_pfr})} &   \multicolumn{1}{c}{Eq.~(\ref{eq_gsr_pfr})} \\\hline
$\Lambda^{\rm{Sr}}_{xxx}$ \T\B  & 97.616     & 111.596     & 97.616 \\
$\Lambda^{\rm{Sr}}_{xyy}$  \T\B & 86.323     & 86.3228     & 86.323 \\
$\Lambda^{\rm{Sr}}_{xxz}$ \T\B  & $-$14.060  & $-$14.060   & $-$14.060 \\    
$\Lambda^{\rm{Sr}}_{xzx}$ \T\B  & $-$14.060  & $-$7.421    & $-$14.060 \\
$\Lambda^{\rm{Ti}}_{yyy}$ \T\B  & 29.021     & 16.800      & 29.021 \\
$\Lambda^{\rm{Ti}}_{yzz}$ \T\B  & $-$360.977 & $-$360.977  & $-$360.977 \\
$\Lambda^{\rm{Ti}}_{yyx}$ \T\B  & $-$18.556  & $-$18.556   & $-$18.556 \\
$\Lambda^{\rm{Ti}}_{yxy}$ \T\B  & $-$18.556  & $-$56.446   & $-$18.556 \\
$\Lambda^{\rm{O}}_{zzz}$ \T\B   & 1887.068   & 2110.762    & 1887.065 \\
$\Lambda^{\rm{O}}_{zxx}$  \T\B  & 168.414    & 168.414     & 168.414 \\
$\Lambda^{\rm{O}}_{zzy}$  \T\B  & 262.311    & 262.309     & 262.309 \\ 
$\Lambda^{\rm{O}}_{zyz}$  \T\B  & 262.311    & 276.029     & 262.309 \\ \hline \hline 
\end{tabular}
\end{center}
\end{table}

In Table~\ref{Tab_STO_forces} we show our calculated CI flexoelectric force-response coefficients, 
and compare them with the ones reported in Ref.~\onlinecite{chapter-15}, where they are directly obtained from the second moments of the FC (as e.g in Eq.~(\ref{eq_sqrbkt})). As evident from the data presented in the table, our results compare quite well with the published ones despite the fact that a different set of pseudopotentials and cell parameters are employed in the two calculations.
In the two bottom rows of Table~\ref{Tab_STO_forces} we also show 
the CI elastic coefficients of c-SrTiO$_3$ calculated via the elastic sum 
rule Eq.~(\ref{eq_sr_celast}) together with the corresponding values obtained with the HWRV calculation. The comparison between the two sets of data evidences, one more time, an excellent agreement.

\begin{table}
\setlength{\tabcolsep}{9pt}
\begin{center}
\caption{\label{Tab_dist_Celast} Selected CI elastic tensor coefficients (in GPa) of a distorted SrTiO$_{3}$ system. Different columns show the elastic tensor as obtained from different approaches (from left to right): sublattice summation of flexoelectric force-response tensor (Eq.~(\ref{eq_sr_celast})), right-hand side of Eq.~(\ref{eq_gsr_celast_pract}), HWRV,~\cite{hamann-metric} and Lagrange as calculated from HWRV (Eq.~(\ref{hwrv_celast})).} %
\begin{tabular}{c|rrrr}\hline\hline
   \T\B   & \multicolumn{1}{c}{$\cal{\bar{C}_{\alpha\gamma,\beta\delta}}$} & \multicolumn{1}{c}{Eq.~(\ref{eq_gsr_celast_pract})} & \multicolumn{1}{c}{$\mathcal{\bar{C}}^{\rm HWRV}_{\alpha\gamma,\beta\delta}$} & \multicolumn{1}{c}{$\bar{C}^{\rm La}_{\alpha\gamma,\beta\delta}$}  \\\hline
$xx,xx$  \T\B   & 219.440 & 219.432 & 227.631 & 211.242\\
$yy,zz$  \T\B   & 164.706 & 164.696 & 113.210 & 113.200 \\
$zz,yy$  \T\B   & 103.990 & 103.981 & 113.209 & 113.200 \\    
$xy,xy$  \T\B   & 120.673 & 120.674 & 120.159 & 111.455 \\
$yx,xy$  \T\B   & 119.645 & 119.645 & 120.159 & 111.455 \\
$yz,yz$  \T\B   & 45.559  & 45.560  & 75.917  &  97.056 \\ 
$yz,zy$  \T\B   & 45.559  & 45.560  & 75.917  &  45.560 \\ \hline \hline 
\end{tabular}
\end{center}
\end{table}

\subsection{Distorted SrTiO$_3$ \label{sec:sto_dist}}

In this section, we numerically test the forces- and stress-mediated \emph{improper} contributions to the sum rules 
derived in Sec.~\ref{sec:elast}. 
To this end, we study a fictitious SrTiO$_3$ lattice where we artificially create finite stresses and forces. 
In particular, we randomly distort in the range of $\pm5\%$ the primitive vectors and the relative atomic coordinates 
from their equilibrium values in the 5-atom primitive cubic cell. 

First, we focus on the piezoelectric force-response tensor by comparing the outcomes of three different calculations: 
i) the HWRV~\cite{hamann-metric} one, which we regard as our reference, 
ii) the sublattice summation of the $\Phi^{(1,\gamma)}_{\kappa\alpha,\kappa'\beta}$ tensor [Eq.~(\ref{eq_sr_pfr})], and 
iii) the same sublattice summation of $\Phi^{(1,\gamma)}_{\kappa\alpha,\kappa'\beta}$, but including the improper 
contribution from the residual forces [Eq.~(\ref{eq_gsr_pfr})]. 
The results shown in Table~\ref{Tab_dist_pfr} nicely confirm our formal prediction: the sublattice 
sum of $\Phi^{(1,\gamma)}_{\kappa\alpha,\kappa'\beta}$ coincides with $\Lambda_{\alpha,\beta\delta}^{\kappa}$
only if the improper contribution is taken into account, otherwise the error can be large, and the expected 
symmetry in the strain indices ($\beta\delta$) is clearly violated.

\begin{table*}
\setlength{\tabcolsep}{10.0pt}
\begin{center}
\caption{\label{Tab_tSTO_flexo} Breakdown of the linearly-independent flexoelectric coefficients (in nC/m) of t-SrTiO$_3$. The rightmost column shows the open-circuit flexovoltages (in V). Data are presented following the same scheme as in Table~\ref{Tab_cubic_flexo}}
\begin{tabular}{c|rrrrr|r}\hline\hline
	\T\B& \multicolumn{1}{c}{$\bm{\bar{\mu}}$} & \multicolumn{1}{c}{$\bm{P}^{(1)}\bm{\Gamma}$} &  \multicolumn{1}{c}{$\bm{\bar{C}}$} & \multicolumn{1}{c}{$\bm{\Phi}^{(1)}\bm{\Gamma}$} & \multicolumn{1}{c}{$\bm{\mu}$} & \multicolumn{1}{c}{$\bm{\varphi}$} \\\hline
 $xx,xx$ \T\B & $-$0.946 & 0.056    &  $-$68.865 &  3.516     &   $-$66.239  & $-$18.086 \\     
 $zz,zz$ \T\B & $-$0.898 & $-$0.013 &  $-$55.841 &  $-$2.507  &   $-$59.259  & $-$18.765 \\ \hline
 $xx,yy$ \T\B & $-$0.786 & 0.052    &  $-$55.977 &  3.416     &   $-$53.294  & $-$14.551 \\    
 $xx,zz$ \T\B & $-$0.830 & $-$0.041 &  $-$58.133 &  $-$2.626  &   $-$61.631  & $-$16.828 \\
 $zz,xx$ \T\B & $-$0.841 & 0.017    &  $-$49.897 &  3.308     &   $-$47.413  & $-$15.014 \\ \hline
 $xy,xy$ \T\B & $-$0.028 & 0.002    &  $-$4.091  &  0.313     &   $-$3.805   & $-$1.039 \\
 $xz,xz$ \T\B & $-$0.079 & 0.023    &  $-$6.796  &  $-$1.292  &   $-$8.144   & $-$2.224 \\
 $zx,xz$ \T\B & $-$0.081 & 0.013    &  $-$5.468  &  $-$1.706  &   $-$7.242   & $-$2.293 \\ \hline \hline
\end{tabular}
\end{center}
\end{table*}

Next, we study (at the CI level) the second-order elastic coefficients of the distorted SrTiO$_3$ crystal 
as obtained from the different strain tensor definitions of Sec.~\ref{sec:elast}, and compare them
to the sublattice 
summation of the CI flexeoelectric force-response tensor [Eq.~(\ref{eq_sr_celast})]. 
In particular, we consider: (i) HWRV~\cite{hamann-metric} ($\mathcal{\bar{C}}^{\rm HWRV}_{\alpha\gamma,\beta\delta}$) (ii) Lagrange ($\bar{C}^{\rm La}_{\alpha\gamma,\beta\delta}$), which we obtain from $\mathcal{\bar{C}}^{\rm HWRV}_{\alpha\gamma,\beta\delta}$ via Eq.~(\ref{hwrv_celast})
(iii) $\cal{\bar{C}_{\alpha\gamma,\beta\delta}}$, given by the sum of $C^\kappa_{\alpha\gamma,\beta\delta}$ [Eq.~(\ref{eq_sr_celast})] (iv) the same sublattice sum of $C^\kappa_{\alpha\gamma,\beta\delta}$, but including the improper
contribution [Eq.~(\ref{eq_gsr_celast_pract})] from the residual stress.
The results are summarized in Table~\ref{Tab_dist_Celast}. As a general observation, it is clear
that the presence of residual stresses breaks the equivalence between the different
versions of the elastic tensor discussed in Sec.~\ref{sec:elast}. 
The data shown in the first column confirm that 
$\cal{\bar{C}_{\alpha\gamma,\beta\delta}}$ is only symmetric with respect to 
the permutation of the second pair of indices ($\beta,\delta$);
on the other hand, the symmetries with respect to ($\alpha, \gamma$) and with respect to an exchange of 
the last two with the first two pair of indices are both broken.
In contrast, the HWRV and Lagrange elastic coefficients, respectively shown in the third and fourth columns, 
manifestly enjoy the whole set of symmetries of a proper elastic tensor. 
Finally, the excellent match between the data of the first two columns 
numerically demonstrates
the generalized elastic sum rule expressed in Eq.~(\ref{eq_gsr_celast_pract}).

\subsection{Tetragonal SrTiO$_3$}

SrTiO$_3$ undergoes, at $T_{\rm_c}\approx105$ K, a cubic-to-tetragonal phase
transition driven by the antiferrodistortive tilts of the oxygen octahedra,
leading to a $I4/mcm$ 
ground state ($a^0a^0c^-$ in Glazer notation).
After the transition, the tilt modes become Raman active, and therefore 
mediate, in principle, an indirect contribution to the flexoelectric tensor 
as discussed in Sec.~\ref{sec:basic}. 
Interestingly, a marked change in the flexoelectric response of SrTiO$_3$ was reported
experimentally~\cite{pavlo} upon cooling the sample below $T_{\rm_c}$, 
which suggests an important contribution of the tilt modes.
From this perspective, tetragonal SrTiO$_3$ (t-SrTiO$_3$) appears as an excellent 
physical example to showcase our methods.

Our calculated values for the flexoelectric coefficients, together with their
decompositions into the contributions described in Sec.~\ref{sec:basic}, are shown in Table~\ref{Tab_tSTO_flexo}.
(As above, we also verified that the sum rules described in the previous paragraphs are 
accurately fulfilled; results are not shown.)
The larger number of independent components (8, compared to only 3 in the $Pm\bar{3}m$ phase) is 
due to the lower symmetry of the tetragonal cell.
The CI coefficients $\bar{\mu}$ show little variation compared to their cubic counterparts, shown in Table~\ref{Tab_cubic_flexo}.
On the other hand, the direct LM contributions are systematically suppressed by about a factor 
of three.
This is simply explained by the stiffening of the lowest-lying transverse polar modes, whose calculated frequencies
are $\omega_{E_u}=78.034$ cm$^{-1}$ and $\omega_{A_{2u}}=87.073$ cm$^{-1}$ for polarization perpendicular ($E_u$)
or parallel ($A_{2u}$) to the tilt axis. 
(As we stated earlier, these modes are primarily responsible for the large LM response in SrTiO$_3$, 
and their contribution to the flexoelectric polarization scales like the inverse square of their
frequency~\cite{sge}.)
Such an interpretation is supported by our analysis of the bulk open-circuit voltage: all type of deformations result in a 
similar value (cf. rightmost columns of Tables~\ref{Tab_cubic_flexo} and~\ref{Tab_tSTO_flexo}) in both the cubic and the tetragonal phase, confirming that the suppression is almost entirely 
due to the smaller dielectric susceptibility of the latter.

As anticipated, we obtain finite indirect contributions to both the electronic and lattice-mediated parts
(see data in second and fourth columns of Table~\ref{Tab_tSTO_flexo}). Although these are comparatively smaller in magnitude than the 
CI contributions, they account for a nonnegligible fraction of the total coefficients.
(Note, e.g., the $\mu_{xz,xz}$ case wherein the indirect contributions amount to $\sim41$\% and $\sim16$\% of the total electronic
and lattice-mediated parts, respectively.)
To understand the origin of such contributions, it is useful to identify the lattice modes
whose coupling to strain is strongest.
A mode decomposition of the internal-strain response tensor $\Gamma$ shows that the 
lattice response to a uniform strain is dominated by tilts, via a nonlinear coupling term
known as \emph{rotostriction}.
A gradient of the tilts, in turn, produces a polarization via the ``rotopolar'' coupling,~\cite{rotopolar}
which is activated in presence of a uniform tilt. 
While promising in light of the observations of Ref.~\onlinecite{pavlo},
a quantitative analysis of our results in terms of these mechanisms goes, however, beyond the 
scopes of the present work, and will be the topic of a forthcoming publication.

\section{Conclusions and outlook}

\label{sec:conclusions}

We have derived the long-wave DFPT~\cite{Royo2019} formulas to calculate the two spatial-dispersion quantities 
necessary to build the lattice-mediated contributions to the bulk flexoelectric tensor. 
These are, the first real-space moment of the interatomic FC matrix and the CI flexoelectric force-response tensor.
We have also generalized to crystals out of mechanical equilibrium
the sum rules that relate the latter quantities
to the established theory of linear elasticity.
We have implemented our formalism in the ABINIT package, thus completing
our earlier developments in the context of the electronic flexoelectric tensor.~\cite{Royo2019} 
We carried out extensive numerical 
tests to benchmark our results against the literature, and demonstrated the performance 
of our method by calculating the bulk flexoelectric tensor of the low-temperature 
tetragonal structure of SrTiO$_3$.

This work culminates a series of theoretical advances carried out during
the last decade on the first-principles theory of flexoelectricity and 
opens the way to the systematic calculation of the bulk flexoelectric tensor 
for any insulating crystal or nanostructure.
We expect this work to greatly facilitate the construction of higher-level 
theories involving gradient-mediated effects.
Indeed, both the first moment of the FC and the flexoelectric 
force-response tensor (via its reformulation as a flexocoupling~\cite{sge})
are the necessary parameters in order to describe, respectively, rotopolar 
and flexolectric couplings within a continuum thermodynamic 
functional~\cite{rotopolar} or an effective Hamiltonian.~\cite{ponomareva}
Surface contributions, albeit not considered here, can be readily incorporated by using our formalism in a 
supercell calculation of a material slab; this has been shown in Ref.~\onlinecite{springolo-21} 
for a set of two-dimensional materials under a flexural deformation.

On the methodological front, work is currently under way to extend our formalism,
both in regards to technical details of the implementation (e.g., for the usability
of the generalized-gradient approximation or nonlinear core corrections in the 
calculation of the flexoelectric coefficients), and in the description of 
related spatial dispersion properties (e.g. natural optical activity or acoustical
activity) based on similar methodologies.
Reports on progress along these lines will be presented in a forthcoming publication.

\begin{acknowledgments}
 We acknowledge the support of Ministerio de Economia,
 Industria y Competitividad (MINECO-Spain) through
 Grant No. PID2019-108573GB-C22  
 and Severo Ochoa FUNFUTURE center of excellence (CEX2019-000917-S);
 and of Generalitat de Catalunya (Grant No. 2017 SGR1506).
 This project has received funding from the European
 Research Council (ERC) under the European Union's
 Horizon 2020 research and innovation program (Grant
 Agreement No. 724529). Part of the calculations were performed at
 the Supercomputing Center of Galicia (CESGA).
\end{acknowledgments}

\appendix

\section{q derivatives of Ewald energy \label{app_ewald}}

The ion-ion Ewald contribution to the FC matrix can be found in Ref.~\onlinecite{gonze-97}. Here, we use a slightly different formulation since we assume the same sublattice-dependent phase factor for the perturbation that was conveniently taken in our previous works (see, e.g., Refs.~\onlinecite{artlin,Royo2019}). The conversion from Ref.~\onlinecite{gonze-97} formula simply consists in multiplying it by $e^{i\mathbf{q}(\boldsymbol{\tau}_{\kappa}'-\boldsymbol{\tau}_{\kappa})}$, which yields
\begin{equation}
\begin{split}
 &E_{{\rm Ew},\mathbf{q}}^{\tau^*_{\kappa\alpha} \tau_{\kappa'\beta}} = Z_{\kappa} Z_{\kappa'}
 \Big[ \frac{4\pi}{\Omega}\sum_{\mathbf{G}}\frac{(G+q)_{\alpha}(G+q)_{\beta}}{(G+q)^2}\,
 e^{i\mathbf{G}(\boldsymbol{\tau}_{\kappa}-\boldsymbol{\tau}_{\kappa'})}\, \\
 & \quad e^{-\frac{(G+q)^2}{4\Lambda^2}} 
   -\sum_{a} \Lambda^3\, e^{i\mathbf{q}{\bf d}_{a,\kappa\kappa'}} \, H^{iso}_{\alpha\beta}(\Lambda{\bf d}_{a,\kappa\kappa'}) \\
 &\quad - \frac{4}{3\sqrt{\pi}}\Lambda^3 \delta_{\kappa\kappa'} 
 e^{i\mathbf{q}(\boldsymbol{\tau}_{\kappa'}-\boldsymbol{\tau}_{\kappa})} \Big].
 \label{eq_ewald}
 \end{split}
\end{equation}
Here $Z_{\kappa}$ is the nuclear charge, ${\bf d}_{a,\kappa\kappa'}={\bf R}_a+\boldsymbol{\tau}_{\kappa'}-\boldsymbol{\tau}_{\kappa}$, $\Lambda$ is a range-separation parameter (setting the Gaussian width) with momentum units that can adopt any value to accelerate the sums convergence, and the specific formula for $H^{iso}_{\alpha\beta}$ is shown in Eq. (25) of Ref.~\onlinecite{gonze-97}. By differentiating the above formula, we arrive at the following {\bf q} gradients of the Ewald contribution (at {\bf q}=0).

The first {\bf q} gradient is given by,
\begin{equation}
\begin{split}
 & E_{{\rm Ew},\gamma}^{\tau^*_{\kappa\alpha}\tau_{\kappa'\beta}}= Z_{\kappa} Z_{\kappa'}
\Bigg\{ \frac{4\pi}{\Omega}  \sum_{{\bf G}} e^{i{\bf G}(\boldsymbol{\tau}_{\kappa}-\boldsymbol{\tau}_{\kappa'})}\, e^{-\frac{G^2}{4\Lambda^2}} \times \\ 
& \quad \Bigg[ \frac{\Big(\delta_{\alpha\gamma}G_{\beta}+\delta_{\beta\gamma}G_{\alpha}\Big)}{G^2} -\frac{2G_{\alpha}G_{\beta}G_{\gamma}}{G^4} -\frac{1}{2\Lambda^2}\frac{G_{\alpha}G_{\beta}G_{\gamma}}{G^2} \Bigg] \\
& \quad -\sum_a \Lambda^3 i \, d_{a,\kappa\kappa',\gamma}\, H^{iso}_{\alpha\beta}(\Lambda{\bf d}_{a,\kappa\kappa'})\Bigg\}.
\label{eq_ewalddq} 
\end{split}
\end{equation}

The second {\bf q} gradient is given by,
\begin{equation}
 \begin{split}
 &  E_{{\rm Ew},\gamma\delta}^{\tau^*_{\kappa\alpha}\tau_{\kappa'\beta},{\rm I}}=
  Z_{\kappa} Z_{\kappa'} \Bigg\{ \frac{4\pi  }{\Omega}  \sum_{{\bf G}}
 e^{i{\bf G}(\boldsymbol{\tau}_{\kappa}-\boldsymbol{\tau}_{\kappa'})}\,
 \frac{e^{-\frac{G^2}{4\Lambda^2}}}{G^2} \times \\
& \quad \Bigg[ \delta_{\alpha\gamma} \delta_{\beta\delta} + \delta_{\beta\gamma} \delta_{\alpha\delta} 
-\Big( \frac{1}{2\Lambda^2} + \frac{2}{G^2} \Big) 
\Big( G_{\alpha}G_{\beta} \delta_{\gamma\delta} \\
& \quad+ G_{\delta} (\delta_{\alpha\gamma}G_{\beta}+\delta_{\beta\gamma}G_{\alpha}) 
 + G_{\gamma} (\delta_{\alpha\delta}G_{\beta}+\delta_{\beta\delta}G_{\alpha})  \Big) \\
 & \quad + \Big( \frac{1}{4 \Lambda^4} +\frac{2 }{\Lambda^2 G^2} + \frac{8}{G^4} \Big) G_{\alpha}G_{\beta} G_{\gamma} G_{\delta} \Bigg] \\
 & \quad +\sum_{a} \Lambda^3\, (d_{a,\kappa\kappa'})_{\gamma}  (d_{a,\kappa\kappa'})_{\delta} \, H^{iso}_{\alpha\beta}(\Lambda{\bf d}_{a,\kappa\kappa'}) \Bigg\} .
 \label{eq_ewalddqdq}
\end{split}
\end{equation}

\section{Second q derivative of atomic displacement Hamiltonian \label{app_Hatdis_d2dqdq}}

The first-order atomic-displacement Hamiltonian consists in a local and a non-local pseudopotential term. The explicit formulas of both contributions, as well as their first {\bf q} derivative (at {\bf q}=0) were reported in our preceeding paper.~\cite{Royo2019} By further differentiation, we arrive at the second {\bf q} gradients of the perturbation, which are necessary to compute the geometric contribution entering the CI flexoelectric force-response tensor (see Eqs.~(\ref{eq_srb_func}) and~(\ref{eq_frozen})).

The local part of the pseudopotential is

\begin{equation}
\begin{split}
 & V_{\gamma\delta}^{{\rm loc},\,\tau_{\kappa\alpha}}({\bf G})= 
  - \frac{i}{\Omega} e^{- i {\bf G} \tau_{\kappa}} \times \\
 & \quad \Bigg( \frac{v_{\kappa}^{\rm loc}(G)'}{G}
  \Big( \delta_{\alpha\delta} G_{\gamma}  
  + \delta_{\alpha\gamma}G_{\delta} + \delta_{\gamma\delta} G_{\alpha} - \frac{G_{\alpha} G_{\delta} G_{\gamma}}{G^2} \Big) \\
  & \quad + \frac{v_{\kappa}^{\rm loc}(G)''}{G^2} G_{\alpha} G_{\delta} G_{\gamma} \Bigg),
\end{split}
\end{equation}
where $G=|{\bf G}|$, $v_{\kappa}^{\rm loc}(G)'$ and $v_{\kappa}^{\rm loc}(G)''$ are the first and second derivatives of the spherical atomic pseudopotential.
The separable part of the pseudopotential is


\begin{equation}
 \begin{split}
   & V^{sep,\tau_{\kappa\alpha}}_{{\bf k},\gamma\delta}({\bf G},{\bf G}')=-\frac{i}{\Omega}\sum_{\mu} e_{\mu\kappa} \, e^{-i({\bf G}-{\bf G}')\boldsymbol{\tau}_{\kappa}} \times \\
   & \qquad \Bigg[\delta_{\alpha\gamma} \zeta_{\mu\kappa,\delta}({\bf k}+{\bf G})\zeta^*_{\mu\kappa}({\bf k}+{\bf G}') \\
   & \qquad + \delta_{\alpha\delta} \zeta_{\mu\kappa,\gamma}({\bf k}  
  +{\bf G})\zeta^*_{\mu\kappa}({\bf k}+{\bf G}') \\
  & \qquad + (k_{\alpha}+G_{\alpha}) \zeta_{\mu\kappa,\gamma\delta}({\bf k}+{\bf G}) \zeta^*_{\mu\kappa}({\bf k}+{\bf G}')  \\
   & \qquad -\zeta_{\mu\kappa,\gamma\delta}({\bf k}+{\bf G})(k_{\alpha}+G_{\alpha}')\zeta^*_{\mu\kappa}({\bf k}+{\bf G}') \Bigg].
 \end{split}
\end{equation}
with $\zeta_{\mu\kappa,\gamma}({\bf k}+{\bf G})$ and $\zeta_{\mu\kappa,\gamma\delta}({\bf k}+{\bf G})$ being the first and second ${\bf q}$ derivatives along the $\gamma$ and $\delta$ Cartesian directions of the separable nonlocal projector.

\section{Phonon--metric stationary functional}

\label{sec:E_st}

At the end of Sec.~\ref{sec:ph-met}, we have shown that the square-bracket tensor can be obtained from the sum of a stationary and a nonvariational second-order functionals,
\begin{equation}
  [\alpha \beta,\gamma \delta]^\kappa = \frac{1}{2} \left( E^{\tau^*_{\kappa\alpha} (\beta)}_{{\rm st},\gamma \delta} + E^{\tau^*_{\kappa\alpha} (\beta)}_{{\rm nv},\gamma \delta} \right).
\end{equation}
Following the arguments of Ref.~\onlinecite{Royo2019}, the stationary part needs to be symmetrized with respect to
$\gamma$ and $\delta$ as 
\begin{equation}
 E^{\tau^*_{\kappa\alpha} (\beta)}_{{\rm st},\gamma \delta}=
 \widetilde{E}^{\tau^*_{\kappa\alpha} (\beta)}_{{\rm st},\gamma \delta}+
 \widetilde{E}^{\tau^*_{\kappa\alpha} (\beta)}_{{\rm st},\delta \gamma},
\end{equation}
where the unsymmetrized (tilded) quantities can be written, using
the relationships between the metric and uniform strain perturbations [Eq.~(\ref{metq1})], as follows
\begin{equation}
\begin{split}
 & \widetilde{E}_{{\rm st},\gamma\delta}^{\tau_{\kappa\alpha}^* (\beta)}= 2 s \int_{{\rm BZ}} [d^3k] \sum_m \widetilde{E}_{{\rm st},m {\bf k},\gamma\delta}^{\tau_{\kappa\alpha}^* (\beta)} \\
 & \quad + i \int_{\Omega} \int K_{\gamma} ({\bf r}, {\bf r}') n^{\tau_{\kappa\alpha}^*}({\bf r}) n^{\eta_{\beta\delta}}({\bf r}') d^3r d^3r', 
 \end{split}
\end{equation}
with a band-resolved contribution that reads
\begin{equation}
\begin{split}
& \widetilde{E}_{{\rm st},m{\bf k},\gamma\delta}^{\tau_{\kappa\alpha}^* (\beta)}= i \langle u_{m{\bf k}}^{\tau_{\kappa\alpha}} | \partial_{\gamma} \hat{H}_{{\bf k}}^{(0)} |   u_{m{\bf k}}^{\eta_{\beta\delta}} \rangle \\
& \quad - i \sum_n \Big( \langle u_{m{\bf k}}^{\tau_{\kappa\alpha}} |u_{n{\bf k}}^{k_{\gamma}} \rangle \langle u_{n{\bf k}}^{(0)} | \mathcal{\hat{H}}_{{\bf k}}^{\eta_{\beta\delta}} | u_{m{\bf k}}^{(0)} \rangle \\
& \qquad \quad \, \, + \langle u_{m{\bf k}}^{(0)} | (\mathcal{\hat{H}}_{{\bf k}}^{\tau_{\kappa\alpha}})^{\dagger} | u_{n{\bf k}}^{(0)} \rangle \langle u_{n{\bf k}}^{k_{\gamma}} | u_{m{\bf k}}^{\eta_{\beta\delta}} \rangle \Big) \\
& \quad + \frac{1}{2} \langle u_{m{\bf k}}^{\tau_{\kappa\alpha}} | \hat{H}_{{\bf k},\gamma\delta}^{(\beta)} | u_{m{\bf k}}^{(0)} \rangle
+ i\langle u_{m{\bf k}}^{(0)} | (\hat{H}_{{\bf k},\gamma}^{\tau_{\kappa\alpha}})^{\dagger} | u_{m{\bf k}}^{\eta_{\beta\delta}} \rangle. 
\label{eq_sqrbkt_bskfunc}
\end{split}
\end{equation}

In Sec.~\ref{sec:results}, we present our results in the form of the type-II CI flexoelectric force-response tensor. 
To this end, in practice one can proceed by i) using the above equations to calculate the square-bracket tensor to subsequently convert it to the type-II force-response tensor via Eq.~(\ref{eq_ci_fxfrt}), or by ii) adopting, from the beginning, the type-II version of the second gradient of the 
metric perturbation introduced in Eq.~(B11) of Ref.~\onlinecite{Royo2019}.
The second route allows to write the formulas yielding directly the CI flexoelectric force-response tensor as follows,
\begin{equation}
  \bar{C}^\kappa_{\alpha \gamma,\beta \delta} =  E^{\tau^*_{\kappa\alpha} (\beta\delta)}_{{\rm st},\gamma} + E^{\tau^*_{\kappa\alpha} (\beta\delta)}_{{\rm nv},\gamma},
  \label{eq_Csupk}
\end{equation}
with a stationary part, 
\begin{equation}
\begin{split}
 & E_{{\rm st},\gamma}^{\tau_{\kappa\alpha}^* (\beta\delta)}= 2 s \int_{{\rm BZ}} [d^3k] \sum_m \widetilde{E}_{{\rm st},m {\bf k},\gamma}^{\tau_{\kappa\alpha}^* (\beta\delta)}  \\
 & \quad + i \int_{\Omega} \int K_{\gamma} ({\bf r}, {\bf r}') n^{\tau_{\kappa\alpha}^*}({\bf r}) n^{\eta_{\beta\delta}}({\bf r}') d^3r d^3r', 
 \end{split}
\end{equation}
\begin{equation}
\begin{split}
& E_{{\rm st},m{\bf k},\gamma}^{\tau_{\kappa\alpha}^* (\beta\delta)}= i \langle u_{m{\bf k}}^{\tau_{\kappa\alpha}} | \partial_{\gamma} \hat{H}_{{\bf k}}^{(0)} |   u_{m{\bf k}}^{\eta_{\beta\delta}} \rangle \\
& \quad - i \sum_n \Big( \langle u_{m{\bf k}}^{\tau_{\kappa\alpha}} |u_{n{\bf k}}^{k_{\gamma}} \rangle \langle u_{n{\bf k}}^{(0)} | \mathcal{\hat{H}}_{{\bf k}}^{\eta_{\beta\delta}} | u_{m{\bf k}}^{(0)} \rangle \\
& \qquad \quad \, \, + \langle u_{m{\bf k}}^{(0)} | (\mathcal{\hat{H}}_{{\bf k}}^{\tau_{\kappa\alpha}})^{\dagger} | u_{n{\bf k}}^{(0)} \rangle \langle u_{n{\bf k}}^{k_{\gamma}} | u_{m{\bf k}}^{\eta_{\beta\delta}} \rangle \Big) \\
& \quad + \frac{1}{2} \langle u_{m{\bf k}}^{\tau_{\kappa\alpha}} | \hat{H}_{{\bf k},\gamma}^{(\beta\delta)} | u_{m{\bf k}}^{(0)} \rangle
+ i\langle u_{m{\bf k}}^{(0)} | (\hat{H}_{{\bf k},\gamma}^{\tau_{\kappa\alpha}})^{\dagger} | u_{m{\bf k}}^{\eta_{\beta\delta}} \rangle. 
\end{split}
\end{equation}
The last three equations have been explicitly used in our code implementation of the CI flexoelectric force-response tensor.~\cite{ABINIT2020}

\section{Treatment of the electrostatic divergence at G=0 \label{app_g0} }

In Ref.~\onlinecite{Royo2019} we exposed how to deal with the divergences at {\bf G}=0 in the cases of the atomic displacement and the metric-wave perturbations, for which finite contributions appeared at first order in {\bf q}. The resulting constant terms affected those matrix elements within the stationary functionals where $\hat{H}^{\tau_{\kappa\alpha}}_{\gamma}$ or $\hat{H}^{\eta_{\beta\delta}}$ appeared bracketed between two unperturbed valence band states. These same {\bf G}=0 contributions need to be taken into account when solving the corresponding stationary functionals for the properties reported in this paper.
Next we discuss how to deal with other two divergences associated with the geometric and the Ewald terms; which become only relevant at second order in {\bf q}, namely, in the calculation of the CI flexoelectric force-response tensor. 

\subsection{Geometric contribution}

The geometric contribution, Eq.~(\ref{deltae}), presents a {\bf G}=0 divergence associated with the local pseudopotential part of the atomic-displacement Hamiltonian. After performing the summation over valence bands and the BZ integration, explicitly appearing in the functionals wherein the geometric term and its gradients participate (see e.g. Eqs.~(\ref{eq_geom0}) and~(\ref{eq_atdismet_nv})), the local pseudopotential part being relevant in the context of a long-wave expansion can be written as,
\begin{equation}
\begin{split}
\Delta E^{{\rm loc},\tau_{\kappa\alpha}^* (\beta)}_{{\bf q}}=
2 \Omega & \int d^3r \left(V_{{\bf q}}^{{\rm loc},\tau_{\kappa\alpha}}({\bf r})\right)^{\dagger} \\ 
& \quad \left(\frac{\partial n^{(0)}({\bf r})}{\partial r_{\beta}} +i \frac{q_{\beta}}{2} n^{(0)}({\bf r})\right),
\end{split}
\end{equation}
where the quantity between round brackets at the second line is minus the geometric contribution to the first-order charge-density response to a metric wave [see e.g. Eq.~(\ref{eq_n_phon_vs_met})]. The above expression is conveniently reformulated in reciprocal space as follows,
\begin{equation}
\begin{split}
\Delta E^{{\rm loc},\tau_{\kappa\alpha}^* (\beta)}_{{\bf q}} &=
2\Omega \sum_{{\bf G}} \left(V_{{\bf q}}^{{\rm loc},\tau_{\kappa\alpha}}({\bf G})\right)^{\dagger} (i G_{\beta} n^{(0)}({\bf G})) ) \\
& \quad + 2\Omega \sum_{{\bf G}} \left(V_{{\bf q}}^{{\rm loc},\tau_{\kappa\alpha}}({\bf G})\right)^{\dagger} (i \frac{q_{\beta}}{2}  n^{(0)}({\bf G})),
\end{split}
\end{equation}
where, after retaining only the {\bf G}=0 term, we obtain,
\begin{equation}
\begin{split}
 \Delta E^{{\rm loc},\tau_{\kappa\alpha}^* (\beta)-0}_{{\bf q}} = &
 \Omega \left(V_{{\bf q}}^{{\rm loc},\tau_{\kappa\alpha}}({\bf G}=0)\right)^{\dagger} \\
 & \qquad  \left(i q_{\beta}  n^{(0)}({\bf G}=0)\right).
 \label{eq_g0_met_loc}
\end{split}
\end{equation}

At this point we proceed following Ref.~\onlinecite{Royo2019} and substitute the divergent contribution of the local potential by its long-wave expansion up to second order in
{\bf q},
\begin{equation}
 V_{{\bf q}}^{\rm loc,\tau_{\kappa \alpha}}({\bf G}=0) \sim  -\frac{i q_{\alpha}}{\Omega}  \left( -\frac{4\pi Z_\kappa}{q^2} + \frac{F_{\kappa}''}{2}  \right).
\end{equation}
Here, $Z_{\kappa}$ is the pseudopotential charge, $q=|{\bf q}|$ and $F_{\kappa}''$ is the second gradient of $F_{\kappa}(q)=q^2 v^{\rm loc}_{\kappa}(q)$.
Pluggin this in Eq.~(\ref{eq_g0_met_loc}) one obtains the following expression for the divergent {\bf G}=0 contribution of the geometric term,
\begin{equation}
\begin{split}
 \Delta E^{{\rm loc},\tau_{\kappa\alpha}^* (\beta)-0}_{{\bf q}}&= 
  -4 \pi Z_{\kappa} n^{(0)}({\bf G}=0) \frac{q_{\alpha}q_{\beta}}{q^2} \\
  & \quad + \frac{1}{2} q_{\alpha}q_{\beta} n^{(0)}({\bf G}=0) F_{\kappa}''(q).
 \label{eq_g0geom}
 \end{split}
 \end{equation}

Recalling that $n^{(0)}({\bf G}=0)=\sum_{\kappa}Z_{\kappa}/\Omega$, the first term at the rhs of the above equation is appropriately incorporated in the ${\bf G}=0$ term of the ionic Ewald contribution (see next section).
The remaining term vanishes for {\bf q}=0 both at zeroth and first order in {\bf q}, however it introduces a finite contribution at second order. After differentiating, we arrive at the ${\bf G}=0$ term to incorporate in the geometric contribution 
to the CI flexoelectric force-response tensor 

\begin{equation}
\Delta E^{\tau_{\kappa\alpha}^* (\beta)-0}_{{\bf \gamma\delta}}= 
 \frac{1}{2} n^{(0)}({\bf G}=0) F_{\kappa}''(q=0) 
\left( \delta_{\alpha\gamma}\delta_{\beta\delta} + \delta_{\alpha\delta} \delta_{\beta\gamma} \right).
\end{equation}

\subsection{Ewald contribution \label{g0_Ew}}

The {\bf G}=0 term of the reciprocal-space sum in the ionic Ewald contribution, Eq.~(\ref{eq_ewald}), can be written as
 \begin{equation}
 E_{Ew,\mathbf{q}}^{\tau^*_{\kappa\alpha} \tau_{\kappa'\beta}-0}= \frac{4 \pi Z_{\kappa} Z_{\kappa'}}{\Omega} q_{\alpha}q_{\beta} \tilde{g}({\bf q}),
 \end{equation}
 with
\begin{equation}
 \tilde{g}({\bf q})=\frac{e^{-\frac{q^2}{4\Lambda^2}}-1}{q^2}.
\label{ewpart}
\end{equation}
In building this equation we have imported the first term of Eq.~(\ref{eq_g0geom}), hence the $-1$ term in the numerator of $\tilde{g}({\bf q})$. This step is crucial to attain a convergent description in the long-wave limit. Indeed, it allows us to expand $\tilde{g}({\bf q})$ up to second order in $q$ as, 
\begin{equation}
 \tilde{g}({\bf q})\sim-\frac{1}{4\Lambda}+\frac{q^2}{32\Lambda^4}.
\end{equation}

Likewise the geometric case, the resulting {\bf G}=0 Ewald term has only a finite contribution at second order in {\bf q} given by,  

\begin{equation}
E_{Ew,\gamma\delta}^{\tau^*_{\kappa\alpha} \tau_{\kappa'\beta}-0}= -\frac{4 \pi Z_{\kappa} Z_{\kappa'}}{\Omega} \left[ \frac{1}{4\Lambda^2}\left( \delta_{\alpha\gamma}\delta_{\beta\delta} + \delta_{\alpha\delta} \delta_{\beta\gamma}\right) \right].
\end{equation}

\section{Real-space representation}

\label{app_realspace}

As an alternative derivation of the sum rules discussed in the main text, 
we shall rewrite the quantities that we have defined in the earlier Sections 
as real-space moments of the FC.
To see how, it is useful to recall the Fourier transforms that
link the FC matrix in real and reciprocal space,
\begin{equation}
\label{phiq}
\Phi_{\kappa \alpha,\kappa' \beta}({\bf q}) = \sum_l \Phi^l_{\kappa \alpha,\kappa' \beta} e^{i {\bf q}\cdot ({\bf R}_{l\kappa'} - {\bf R}_{0\kappa})}.
\end{equation}
Here, $l$ is a cell index, the real-space vectors ${\bf R}_{l\kappa}={\bf R}_{l}+{\bm \tau}_{\kappa}$ span the crystal lattice, and
\begin{equation}
\label{phil}
\Phi^l_{\kappa \alpha,\kappa' \beta} = \frac{\partial^2 E}{\partial R_{0\kappa \alpha} \partial R_{l\kappa' \beta}},
\end{equation}
defines the interatomic force constants as second derivatives of the total energy
with respect to atomic displacements.
By differentiating with respect to ${\bf q}$ we find
\begin{subequations}
\label{realmom}
\begin{align}
\Phi_{\kappa \alpha,\kappa' \beta}^{(1,\gamma)}       = & \sum_l \Phi^l_{\kappa \alpha,\kappa' \beta}
      ({\bf R}_{0\kappa}- {\bf R}_{l\kappa'})_\gamma, \\
\Phi_{\kappa \alpha,\kappa' \beta}^{(2,\gamma \delta)} = & \sum_l \Phi^l_{\kappa \alpha,\kappa' \beta} 
 ({\bf R}_{0\kappa}- {\bf R}_{l\kappa'})_\gamma ({\bf R}_{0\kappa}- {\bf R}_{l\kappa'})_\delta.
\end{align}
\end{subequations}
Note that, in real crystals, the interatomic forces are always 
long-ranged, and the above lattice sums are conditionally
convergent. 
Thus, before performing the sums we shall remove the
long-range part of the FC via either a fictitious
Thomas-Fermi gas,~\cite{martin,artlin} or an Ewald-like Gaussian filter.~\cite{Royo2021}
Generally the results do depend on the arbitrary length scale, $\sigma$,
that we use to perform the range separation; however, one can
show that in the $\sigma \rightarrow 0$ limit we recover 
our definition of the analytic ${\bf q}$-derivatives 
within the assumption of short-circuit EBC.

Based on the above, we can make our progress towards the sum rules by 
considering a bounded crystallite, consisting in a large number $N$ of 
crystal cells.
We assume that the volume is large enough so the surface-specific 
deviations in the interatomic force constants and/or site occupancies
is irrelevant for what follows; for the same reason, we shall use the same 
bulk-like labeling of the atomic sites in terms of a combined cell ($l$)
and sublattice ($\kappa$) index.
Since the lattice is no longer periodic, however, we need to consider 
the full force constants matrix of the crystallite, in the form
\begin{equation}
\label{phill}
\Phi^{ll'}_{\kappa \alpha,\kappa' \beta} = \frac{\partial^2 E}{\partial R_{l\kappa \alpha} \partial R_{l' \kappa' \beta}}.
\end{equation}
If we suppose that the $l=0$ cell is located far away from the surfaces, 
and that the force constants decay fast enough with distance, we can 
readily identify $\Phi^{0l'}_{\kappa \alpha,\kappa' \beta}$ of Eq.~(\ref{phill})
with $\Phi^{l'}_{\kappa \alpha,\kappa' \beta}$ of Eq.~(\ref{phil}),
which leads to a straightforward redefinition of the real-space moments of Eq.~(\ref{realmom}).
Since the surface to bulk volume ratio is assumed to be negligible, we 
eventually arrive
at the following lattice sums,
\begin{subequations}
\label{phi12}
\begin{align}
\sum_{\kappa'} \Phi_{\kappa \alpha,\kappa' \beta}^{(1,\gamma)}       = & -\sum_{ l' \kappa'} 
   \Phi^{0l'}_{\kappa \alpha,\kappa' \beta} ({\bf R}_{l'\kappa'})_\gamma,\\
\sum_{\kappa \kappa'} \Phi_{\kappa \alpha,\kappa' \beta}^{(2,\gamma \delta)} = & - \frac{1}{N} \sum_{l\kappa l' \kappa'} 
  \Phi^{ll'}_{\kappa \alpha,\kappa' \beta} \times \nonumber \\
 & \left[ ({\bf R}_{l\kappa})_\gamma({\bf R}_{l'\kappa'})_\delta + ({\bf R}_{l\kappa})_\delta  ({\bf R}_{l'\kappa'})_\gamma \right], \label{phi2}
\end{align}
\end{subequations}
In deriving the above formulas we have used the translational invariance
of the whole crystallite, 
\begin{equation}
\sum_{l'\kappa'} 
   \Phi^{ll'}_{\kappa \alpha,\kappa' \beta} = \sum_{l\kappa} 
   \Phi^{ll'}_{\kappa \alpha,\kappa' \beta} = 0
\end{equation}   
in order to eliminate the remaining terms.

The first- and second-order coefficients of Eq.~(\ref{eofu}) can be expressed, respectively, 
in terms of the atomic forces and interatomic force constants via Eq.~(\ref{unsymm}).
In particular, by applying the chain rule to the strain derivatives in the second-order terms of
Eq.~(\ref{eofu}) we obtain (we indicate as $E^{\rm cr}$ the total energy of the bounded crystallite
to distinguish it from the bulk energy per cell, $E$)
\begin{subequations}
\begin{align}
\frac{\partial^2 E}{\partial \tau_{\kappa \alpha} \partial u_{\beta\delta}} =& 
 \sum_{l'\kappa'} \frac{\partial^2 E^{\rm cr}}{\partial R_{0\kappa \alpha} \partial R_{l\kappa' \beta}} R_{l\kappa'\delta}, \\ 
\frac{\partial^2 E}{\partial u_{\alpha \gamma}u_{\beta \delta}} =& \frac{1}{N}
  \sum_{l \kappa l' \kappa'} \frac{\partial^2 E^{\rm cr}}{\partial R_{l\kappa \alpha}R_{l'\kappa' \beta}} R_{l\kappa\gamma} R_{l'\kappa'\delta}.
\end{align}
\end{subequations} 
By recalling the definition of the interatomic force constants, Eq.~(\ref{phill}), we can 
directly relate this result to Eq.~(\ref{phi12}), and thereby arive at the sum rules of Eqs.~(\ref{eq_gsr_pfr}) and~(\ref{eq_gsr_celast}).

\bibliography{merged}

\end{document}